\documentclass[a4paper,fleqn,usenatbib]{mnras}

\usepackage{graphicx}
\usepackage{amsmath}
\usepackage{amssymb}
\usepackage{txfonts}

\usepackage[T1]{fontenc}
\usepackage{ae,aecompl}

\usepackage[dvipsnames]{xcolor}
\usepackage{lipsum}
\usepackage{natbib}
\usepackage{multirow}
\usepackage{booktabs}

\usepackage[colorinlistoftodos,prependcaption,textsize=small]{todonotes}

\usepackage{soul}

\title[Structural and size evolution of H$\alpha$-selected SFGs]{The structural and size evolution of star-forming galaxies over the last 11 Gyrs}

\author[A. Paulino-Afonso et al.]{
Ana Paulino-Afonso$^{1,2}$,\thanks{E-mail: aafonso@oal.ul.pt}
David Sobral$^{3,4}$,
Fernando Buitrago$^{1,2}$,
Jos\'e Afonso$^{1,2}$
\\
$^{1}$Instituto de Astrof\'isica e Ci\^encias do Espa\c{c}o, Universidade de Lisboa, OAL, Tapada da Ajuda, PT1349-018 Lisboa, Portugal\\
$^{2}$Departamento de F\'isica, Faculdade de Ci\^encias, Universidade de Lisboa, Edif\'icio C8, Campo Grande, PT1749-016 Lisboa, Portugal\\
$^{3}$Department of Physics, Lancaster University, Lancaster, LA1 4YB, UK\\
$^{4}$Leiden Observatory, Leiden University, P.O. Box 9513, NL-2300 RA Leiden, The Netherlands\\
}

\date{Accepted 2016 November 10. Received 2016 November 10; in original form 2016 July 5}

\pubyear{2016}

\begin{document}

\label{firstpage}
\pagerange{\pageref{firstpage}--\pageref{lastpage}}
\maketitle

\begin{abstract}
{We present new results on the evolution of rest-frame blue/UV sizes and S\'ersic indices of H$\alpha$-selected star-forming galaxies over the last 11 Gyrs. We investigate how the perceived evolution can be affected by a range of biases and systematics such as cosmological dimming and resolution effects. We use \textsc{GALFIT} and an artificial redshifting technique, which includes the luminosity evolution of H$\alpha$-selected galaxies, to quantify the change on the measured structural parameters with redshift. We find typical sizes of 2 to 3 kpc and S\'ersic indices of $n\sim1.2$, close to pure exponential disks all the way from $z=2.23$ to $z=0.4$. At $z=0$ we find typical sizes of $4-5$ kpc. Our results show that, when using GALFIT, cosmological dimming has a negligible impact on the derived effective radius for galaxies with $<10$ kpc, but we find a {$\sim20\%$} bias on the estimate of the median S\'ersic indices, rendering galaxies more disk-like. Star-forming galaxies have grown on average by a factor of {$2-3$} in the last 11 Gyrs with {$r_e\propto(1+z)^{-0.75}$}. By exploring the evolution of the stellar mass-size relation we find evidence for a stronger size evolution of the most massive star-forming galaxies since $z\sim2$, as they grow faster towards $z\sim0$ when compared to the lower stellar mass counterparts. As we are tracing the rest-frame blue/UV, we are likely witnessing the growth of disks where star formation is ongoing in galaxies while their profiles remain close to exponential disks, $n\lesssim1.5$, across the same period.}
\end{abstract}



\bibpunct[ ]{(}{)}{;}{a}{}{,}


\section{Introduction}\label{section:intro}

Ever since the first classification schemes based on the visual appearance of galaxies were created \citep[e.g.][]{hubble1926,vaucouleurs1959}, the ways we study galaxy morphology have evolved dramatically. On one hand, we have improved on the quantification of the light distribution in galaxies either using parametric surface brightness profiles \citep[e.g.][]{caon1993,simard1998,peng2002,peng2010} or non-parametric approaches \citep[e.g.][]{abraham1994,conselice2003,lotz2004,law2007,freeman2013}. On the other hand, there was the need to create new classifications as galaxies become more and more irregular towards higher redshift \citep[e.g.][]{serrano2010,huertas-company2015}. 

Although the observed morphology may not be directly linked with intrinsic properties of the stellar populations and dust/gas content of galaxies \citep[see e.g.][and references therein]{conselice2014}, early studies have shown that rest-frame optical morphology correlates with color and star-formation activity \citep[e.g.][]{holmberg1958} and there is a marked difference in the prevalence of different morphological populations in different environments \citep[e.g.][]{dressler1980}. These effects are seen both in the local \citep[e.g.][]{ball2008,bamford2009} and in the higher redshift Universe \citep[e.g.][]{perez-gonzalez2008,viero2012,basset2013}. Additionally, there was significant work regarding correlations between the shape of a galaxy and other physical properties such as stellar populations, mass and star formation \citep[e.g.][]{ roberts1994,conselice2003, blanton2003,wuyts2011,whitaker2015}. Processes such as galaxy mergers, \emph{in-situ} star formation and accretion of inter-galactic gas can be revealed by a detailed structural analysis of galaxy samples.

The peak of star formation in the Universe occurred at $z\sim2$ \citep[$\sim11$ Gyrs ago, e.g.][and references therein]{madau2014} and it is also since this peak of activity that most of the structures (disk and spheroidal galaxies) that we observe today have been formed \citep[e.g.][]{buitrago2013,mortlock2013}. However, to understand how the baryonic structures grew between different cosmological epochs it is not only necessary to study morphology, but also couple that with kinematic information.

In the recent years, it has been possible to study the interplay between galaxy morphology and kinematics by making use of the new available Integral Field Unit (IFU) instruments. These allowed for large galaxy surveys either in the local Universe (e.g. {CALIFA \citet{sanchez2012}, SAMI \citet{bryant2015}; MaNGA \citet{bundy2015}}) and at high redshift (e.g. {KROSS \citet{stott2016}, KMOS$^\mathrm{3D}$ \citet{wisnioski2015}}), and added valuable information that will provide key insights on the physics that drive galaxy evolution. 

Despite the large potential for progress, to connect observed properties across a large span of time we need to account for biases and systematics which can arise either due to selection or instrumental and/or cosmological effects. To overcome these problems we need both large and homogeneous surveys at various cosmic epochs to minimize the impact of cosmic variance and to probe a wide range of galaxy properties and environments. This is now possible when using surveys based on well known and calibrated physical properties over a wide range of redshifts, such as H$\alpha$ in narrow band-surveys \citep[see e.g.][]{moorwood2000,geach2008,villar2008,sobral2009,ly2011,sobral2011} and in spectroscopic/grism surveys \citep[see e.g.][]{mccarthy1999,yan1999,hopkins2000,shim2009,atek2010,vandokkum2011}. 

Nonetheless, even with an ideal, homogeneous sample, we still need to account for all effects that have a dependence on redshift as they could mimic and/or influence evolutionary trends that we observe. One of the strongest effects that impacts the study of galaxy morphology is the surface brightness dimming \citep{tolman1930}. Between $z\sim0$ and $z\sim2$ this effect impacts the observed fluxes by two orders of magnitude. The first attempts to describe the impact of the surface brightness dimming on how galaxies would be observed at high redshift if they were as we see them today were conducted by \citet{weedman1985}. Later on, studies on the impact on galaxy visual morphology \citep[e.g.][]{giavalisco1996,hibbard1997} and on the morphology quantifiers \citep[e.g.][]{trujillo2007,barden2008,petty2009,weinzirl2011,mosleh2013} were also carried out and find no systematics and errors $\lesssim15\%$ on the derived sizes at $z\sim1$. Nonetheless, these studies are often limited to small samples and comparison between two distinct epochs (one local and one at high redshift).

There are numerous studies reporting on size and structural evolution of galaxies \citep[e.g.][]{ferguson2004,ravindranath2004,trujillo2006,buitrago2008,cimatti2008,franx2008,vanDokkum2010,wuyts2011,mosleh2011,ichikawa2012}. For star-forming galaxies (SFGs) in the range $0.5<z\lesssim3$ studies find moderate \citep[e.g. $r_e\propto(1+z)^{\alpha}$; ][with $\alpha=-0.82,-0.75,-0.57$, respectively]{buitrago2008,vanderwel2014,morishita2014} to negligible size evolution \citep{stott2013b} and light profiles close to exponential disks \citep[e.g.][]{morishita2014,shibuya2015}. For $2<z<4.5$, it was shown that the trend on size evolution (measured from rest-frame UV) depends on the method used to derive galaxy sizes \citep{ribeiro2016}. And, at even higher redshifts ($4<z<8$), \citet{curtis-lake2016}, using FUV rest-frame galaxy sizes, show that the derived evolution depends on the statistical estimators one uses.
These evolutionary trends of galaxy growth and the relation of sizes with stellar masses are also found in large scale cosmological simulations \citep[e.g.][]{genel2014,furlong2015}. However, we do not know yet if such differences can be explained by different selection methods for the definition of the SFGs samples \citep[e.g.][]{oteo2015}. 

Although there are existing studies on the morphologies of SFGs \citep[e.g.][]{vanderwel2014,morishita2014,shibuya2015} and on the quantification of systematic differences of structural properties at low and high redshift \citep[e.g.][]{barden2008,petty2009,weinzirl2011,mosleh2013}, it is still unclear what the role of potential biases and systematics may be. In this paper, we take advantage of a unique H$\alpha$ selection, along with the largest IFU samples, to make further progress on the current open questions.

This paper is organized as follows. In Section \ref{section:sample} we describe the samples of SFGs that will be used throughout the paper. We present our methodology to simulate galaxies at high redshift and to study their structural parameters in Sections \ref{section:artred} and \ref{section:galfit}. The results obtained for our low and high redshift SFGs and simulations are reported in Section \ref{section:results}. We discuss the implications of our results in the context of galaxy evolution in the last $\sim$11 Gyrs in Section \ref{section:discussion}. Finally, in Section \ref{section:conclusions} we summarize our conclusions. Magnitudes are given in the AB system \citep{oke1983}. All the results assume a $\Lambda$CDM cosmological model with $H_0$$=$70.0$\ \mathrm{km\ s^{-1}Mpc^{-1}}$,$\ \Omega_m$$=$0.3 and $\Omega_\Lambda$$=$0.7 and a \citet{chabrier2003} initial mass function.


\section{Galaxy samples}\label{section:sample}

\begin{table*}
\centering
\begin{tabular}{ |c|c|c|c|c|c|c|c| }
 \hline
 & Sample & $\tilde z $ & $N_\mathrm{tot}[N_\mathrm{subsample}]$ & $ \log_{10}(M_\star/M_\odot) $ & $ \log_{10}(L_{\mathrm{H\alpha}}) $  & $ L_{\mathrm{H\alpha}}/L^*_{\mathrm{H\alpha}}(z)$ \\ \hline
\multirow{4}{*}{\rotatebox[origin=c]{90}{Local}}&  CALIFA & $0.015\ [0.016]$ & 541\ [137] & $10.44\ [10.53]$ & $41.08\ [41.22]$ & $0.16\ [0.22]$\\
& SAMI & $0.039\ [0.049]$ & 2349\ [422] & $10.06\ [10.26]$ & $40.38\ [41.22]$ & $0.03\ [0.21]$\\
& MaNGA & $0.030\ [0.037]$ & 8492\ [1536] & $9.85\ [9.91]$ & $40.25\ [41.14]$ & $0.02\ [0.18]$\\
& NYU-VAGC & $0.041\ [0.041]$ & 1285\ [412] & $9.73\ [10.08]$ & $40.74\ [41.05]$ & $0.07\ [0.14]$\\
\hline
\multirow{4}{*}{\rotatebox[origin=c]{90}{HiZELS}}& NB921 & $0.400$ & 460\ [33] & $8.25\ [9.83]$ & $40.39\ [41.31]$ & $0.02\ [0.18]$\\
& NBJ & $0.840$ & 425\ [309] & $9.69\ [9.96]$ & $41.43\ [41.45]$ & $0.15\ [0.16]$\\
& NBH & $1.470$ & 313\ [250] & $9.65\ [9.89]$ & $42.11\ [42.12]$ & $0.38\ [0.38]$\\
& NBK & $2.230$ & 572\ [526] & $9.71\ [9.74]$ & $42.19\ [42.19]$ & $0.21\ [0.21]$\\
\hline
\end{tabular}
\caption{Samples studied in this paper. The median redshift, median stellar mass and median observed aperture corrected (for SAMI, MaNGA and NYU-VAGC) H$\alpha$ luminosity (in $\mathrm{ergs\ s^{-1}}$) for all samples. The last column shows the ratio of the median observed aperture corrected (for SAMI, MaNGA and NYU-VAGC) H$\alpha$ luminosity to the $L^*_{\mathrm{H\alpha}}$ at the median redshift. The value in brackets shows the median values for each subsample after applying the selection criteria defined in Section \ref{section:sample}.}
\label{tab:numbers}
\end{table*}

The main sample of SFGs that we use in this paper comes from the High-Z Emission Line Survey \citep[HiZELS,][]{sobral2013}. Being the largest H$\alpha$ narrow-band survey at high-redshift it provided targets to be observed with IFU instruments such as VLT/SINFONI { \citep{swinbank2012a,swinbank2012b} and VLT/KMOS {\citep{sobral2013b,stott2014,stott2016}}}. Other samples are selected from the currently on-going IFU surveys in the local Universe (see Section \ref{ssection:lowz}). However, local samples differ from a simple H$\alpha$ selection as done in HiZELS. To ensure that we are studying comparable populations, we apply simple sample selection based on stellar mass and H$\alpha$ luminosity:
\begin{equation}
\log_{10}(M_\star/M_\odot)>9   ~~~ \wedge ~~~  L_\mathrm{H\alpha}>0.1L_\mathrm{H\alpha}^*(z),
\label{eq:samplesel}
\end{equation}
where the luminosity cut is taken from the equation for the redshift evolution of $L_\mathrm{H\alpha}^*$ derived by \citet{sobral2013}
\begin{equation}
\log_{10}(L_\mathrm{H\alpha}^*(z)) = 0.45z +  41.87
\end{equation}
We choose to focus our study by selecting samples through their H$\alpha$ luminosities since such samples should be representative of the full star-forming population \citep[e.g.][]{oteo2015}. Note that for the local Universe samples, with the exception of the CALIFA survey, we only have available H$\alpha$ luminosities measured inside a $3"$ fiber. We thus apply aperture corrections following e.g. \citet[][]{garn2010} by computing the flux ratio of the total and fiber magnitudes in the Sloan Digital Sky Survey (SDSS) $r$-band (the filter that contains H$\alpha$) and applying that correction to the observed fiber-based H$\alpha$ luminosity. These correction factors vary from $\sim1.5$ up to $\sim 40$. Our samples and the selection criteria are shown in Figure \ref{fig:ALL_MassHa}. For a quick summary we refer the reader to Table \ref{tab:numbers}.

\subsection{The low-redshift samples}\label{ssection:lowz}

\subsubsection{The CALIFA sample}

The Calar Alto Legacy Integral Field spectroscopy Area \citep[CALIFA, ][]{sanchez2012} survey is a program conducted using the Potsdam Multi-Aperture Spectrophotometer with PMAS fiber PAcK (PMAS/PPAK) integral field spectrophotometer mounted on the 3.5m Calar Alto telescope. The survey aims at observing $\sim600$ galaxies in the local Universe (redshift range $0.005<z<0.03$), which are selected from the SDSS as a sample limited in apparent diameter ($45"<D<80"$). This constraint assures that galaxies are observed within the large field of view ($\sim1$ arcmin$^2$) with a large covering fraction and high spectral resolution.

We use the reported values by \citet{catalan-torrecilla2015} for the H$\alpha$ luminosities of this sample, which are available for 270 galaxies. By taking the selection criteria defined in equation \ref{eq:samplesel} we reach a final sample of 137 CALIFA galaxies (see Figure \ref{fig:ALL_MassHa}).

\subsubsection{The SAMI target sample}

The Sydney-AAO Multi-object Integral field spectrograph \citep[SAMI, ][]{bryant2015} Galaxy survey proposes to target $3400$ galaxies with the SAMI instrument mounted on the Anglo-Australian Telescope (AAT). This survey targets galaxies in the redshift range $0.004<z<0.095$, SDSS magnitudes $r_\mathrm{pet}<19.4$, from low ($10^{7}M_\odot$) to high stellar mass ($10^{12}M_\odot$), both isolated, in groups or members of clusters with halo masses of $\sim10^{15}M_\odot$. Most of the targets (with the exception of few cluster objects) have available SDSS coverage and are selected from the Galaxy and Mass Assembly survey \citep[GAMA, ][]{driver2009}. 

\begin{figure}
\centering
\includegraphics[width=\linewidth]{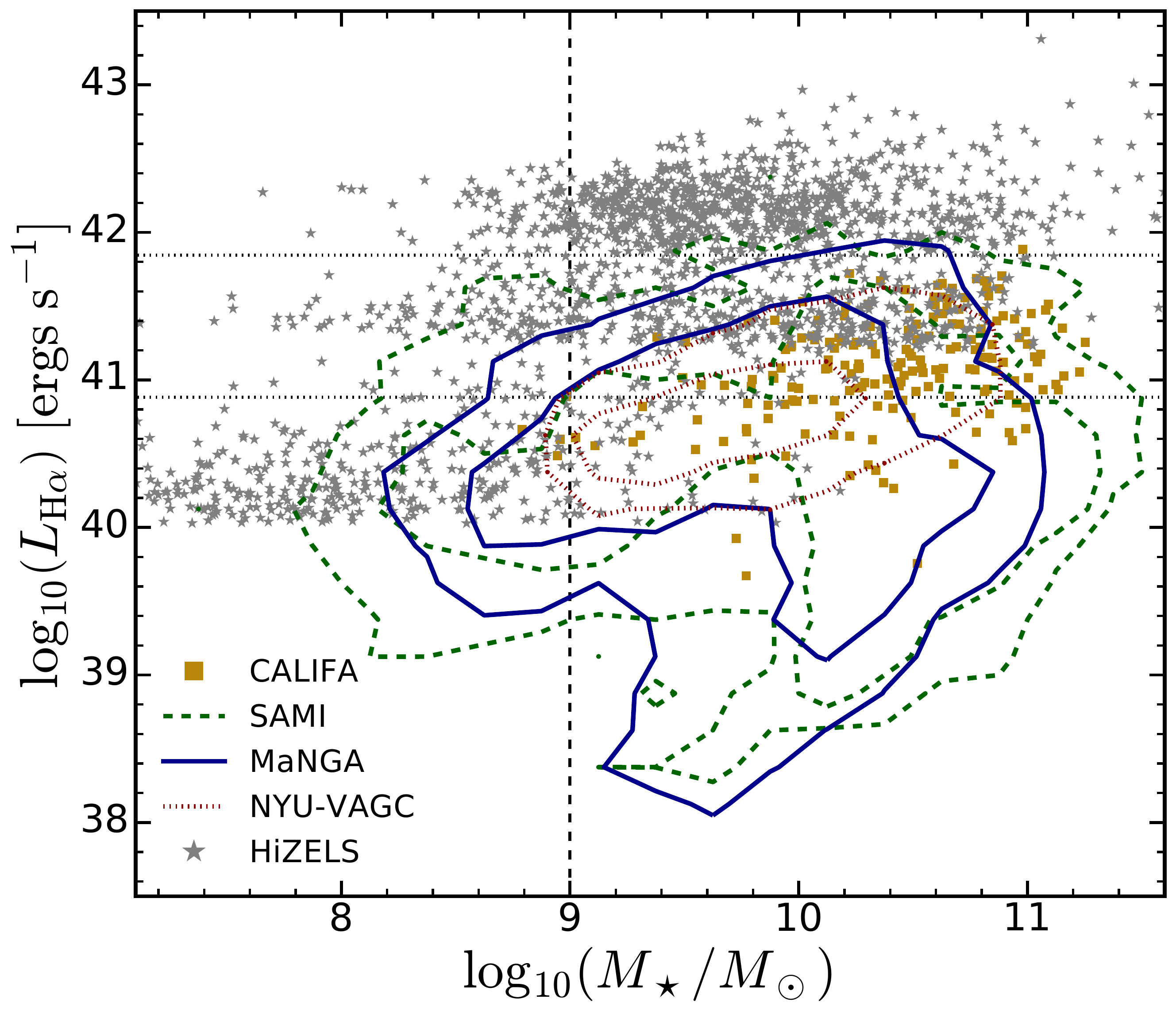}
\caption{The stellar masses and aperture corrected observed H$\alpha$ luminosities of the parent samples used in this paper. The contour lines for SAMI, MaNGA and NYU-VAGC show the limits that contain 68\% and 95\% of the sample for two-dimensional histograms with 0.25 width bins in both stellar mass and H$\alpha$ luminosity. The vertical dashed line shows the stellar mass cut used for the final sample selection. We see that most our samples are above the stellar mass limit imposed in our selection and that the local redshift samples overlap. The horizontal dotted lines show the H$\alpha$ limits at $z=0.01$ (lower line) and at $z=2.23$ (upper line). This shows that the H$\alpha$ luminosity selection as a great impact on the final samples that we study.}
\label{fig:ALL_MassHa}
\end{figure}

We use the GAMA Data Release 2 \citep{liske2015} to obtain H$\alpha$ luminosities for the SAMI sample. By taking the selection criteria defined in equation \ref{eq:samplesel} we reach a final sample of 422 SAMI galaxies (see Figure \ref{fig:ALL_MassHa}). We note that the SAMI sample is stellar-mass complete for our mass limit  only at $z<0.5$. We  include higher redshift galaxies but do not expect it to have a great impact on our derived results.

\subsubsection{A MaNGA-like sample}

The Mapping Nearby Galaxies at Apache point observatory \citep[MaNGA, ][]{bundy2015} survey is part of the SDSS-IV program and aims to study kinematics and internal composition of a sample of $\sim10 000$ galaxies. It will do so with fiber-bundle IFUs with diameters ranging from $12\arcsec$ to $32\arcsec$ and will provide spectral information in the full optical range. The MaNGA sample is derived from an extended version of the NASA Sloan-Atlas (NSA), based on the SDSS DR7 Main Galaxy Sample \citep{abazajian2009}  with the additions and improvements detailed by \citet{blanton2011}\footnotemark{}\footnotetext{\url{http://www.nsatlas.org}}. It will observe galaxies at redshifts $0.01<z<0.15$ with stellar masses above $\sim10^{9}M_\odot$ and it will make use of redshift and $i$-band luminosity to achieve a homogeneous radial coverage \citep[see Figure 8 of][]{bundy2015}, flat stellar mass distributions and a diversity of environments. An additional selection on color space will enhance the targeting of rarer galaxies (green valley, low mass red, and massive blue galaxies).

Using the available data from NSA we attempt to mimic the MaNGA selection by applying the selection bands as described in \citet{bundy2015}. We use the published version of the NSA table (restricting our galaxies to $z<0.055$) and pre-select all galaxies that fall inside the selection bands of Figure 8 by \citet{bundy2015}. We randomly select $\sim6000$ galaxies (with uniform sampling up to 1.5$r_e$, primary selection criteria) and $\sim2500$ galaxies (with uniform sampling up to 2.5$r_e$, secondary selection criteria) to roughly match the sample numbers of MaNGA \citep{bundy2015}. We neglect the color enhanced selection. By taking the selection criteria defined in equation \ref{eq:samplesel} we reach a final sample of 1536 MaNGA galaxies (see Figure \ref{fig:ALL_MassHa}).

\subsubsection{The NYU-VAGC sample}

This sample is based on the New-York University Value Added Galaxy Catalog \citep[NYU-VAGC,][]{blanton2005}. A subset of the catalog, which was constructed as a volume limited sample with well defined selection criteria \citep[see][]{blanton2004}, was chosen as a control sample so that we may inspect if the IFU samples are biased against a magnitude selected sample. To complement the information we matched the catalog with the Max Planck for Astronomy \& Johns Hopkins University Data Release 7 catalogs (MPA-JHU DR7) \citep{kauffmann2003,tremonti2004,salim2007}, which have spectroscopic information for SDSS DR7 galaxies \citep{abazajian2009}.

We first do a pre-selection of SFGs with stellar masses $9<\log_{10}(M_\star/M_\odot)<12$ and H$\alpha$ luminosities greater than $L_\mathrm{H\alpha}>10^{39.5}\mathrm{ergs\ s^{-1}}$ to exclude both low and high stellar masses and match the HiZELS detection limits. This results in a total of $\sim13 000$ galaxies. From this sample, we have randomly selected $10\%$ ($1285$) of all galaxies. This selection was performed by randomly picking galaxies with a probability matched to the magnitude, radius and S\'ersic indices distribution available from the NYU-VAGC catalog. With this method we guarantee that we probe the full morphological parameter space using this subset. 

We then restrict our sample to 412 NYU-VAGC galaxies (see Figure \ref{fig:ALL_MassHa}) with aperture corrected H$\alpha$ luminosities and stellar masses matching our sample selection criteria defined in equation \ref{eq:samplesel}.

\subsection{The high-redshift Universe}\label{ssection:highz}

By using four narrow-band filters in the $z$, $J$, $H$ and $K$-bands, {\citep[HiZELS, ][]{sobral2013}} has detected thousands of H$\alpha$ emitters in four distinct redshifts intervals centered at $0.4$, $0.84$, $1.47$ and $2.23$. Such large samples, selected in an uniform way, allow one to probe galactic evolution across these slices of our Universe that map the decline since the peak of star formation activity. To ensure that the selection is effective in picking up H$\alpha$ emitters, the observations were conducted on well studied extragalactic fields (UDS, COSMOS, SA22, ELAIS N1, Bo\"otes and Lockman Hole), where the ancillary broad band data helps at pinpointing H$\alpha$ emitters by means of color selections and photometric redshifts to allow for the control of the contamination rates \citep{sobral2013}. 

In this paper we make use of the list of emitters that are found in the COSMOS field \citep{sobral2013} for which we have extensive coverage of F814W imaging obtained with the Hubble Space Telescope \citep{scoville2007}. After applying the selection cuts defined in equation \ref{eq:samplesel} we reach final samples with 33 galaxies at $z=0.4$, 309 galaxies at $z=0.84$, 250 galaxies at $z=1.47$ and 526 galaxies at $z=2.23$ (see Table \ref{tab:numbers}).


\section{Artificial redshifting galaxies}\label{section:artred}

\begin{figure*}
\centering
\includegraphics[width=\linewidth]{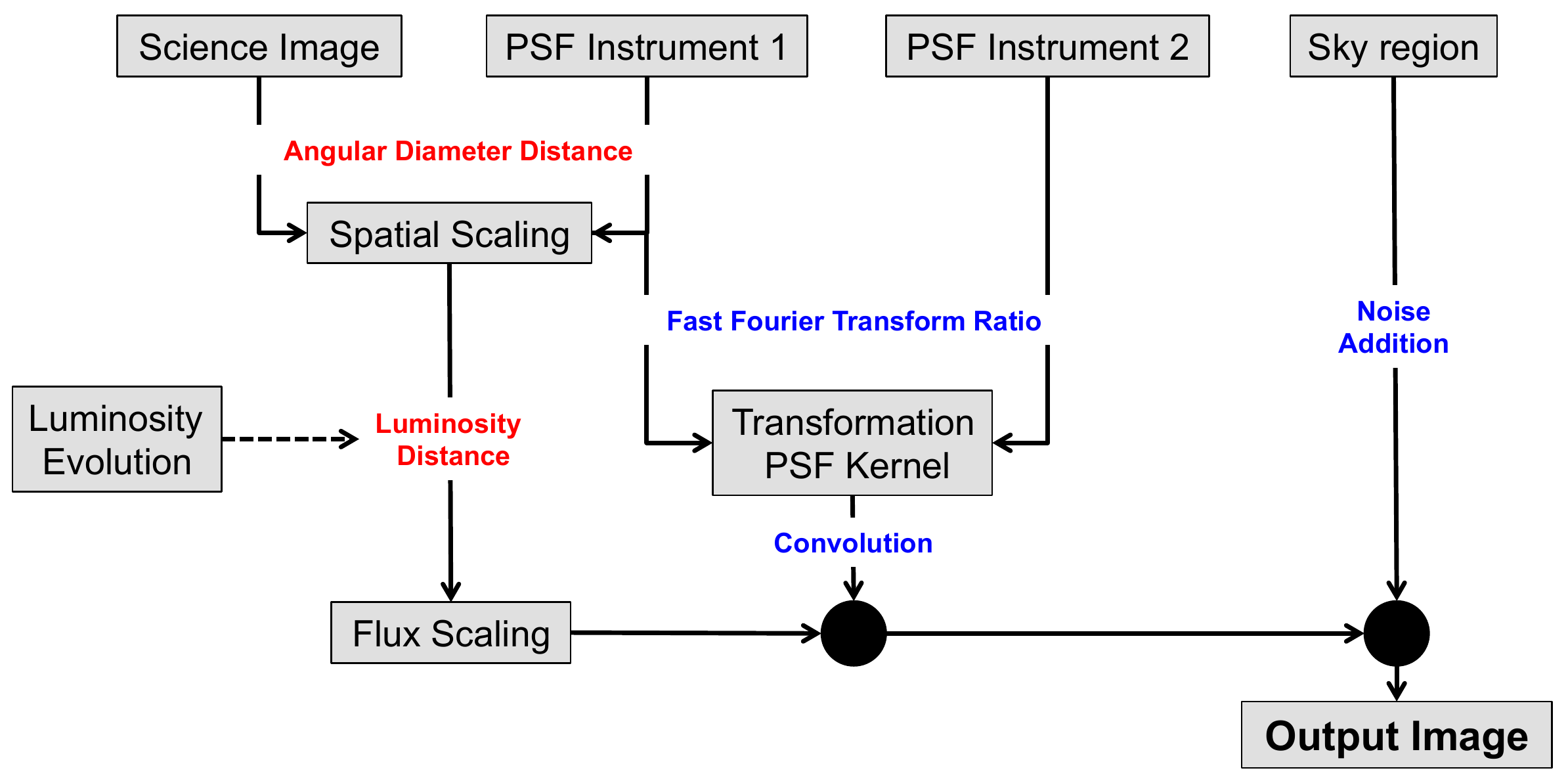}
\caption{Schematic diagram of the algorithm for artificially redshift galaxies. Each step is described in Section \ref{section:artred}.}
\label{fig:dopterian_diagram}
\end{figure*}

To understand how the perceived structural parameters have changed across cosmic time we study how our local galaxies would look like if they were at high redshift. We explore and evaluate the effects of cosmological dimming on various properties of galaxies (e.g. size/shape measurements) taking into account that these are prone to resolution and bandpass issues. 

By artificially redshifting galaxy images, we aim to address the biases and systematics of the parent sample and to build a coherent vision of the galaxies' structural evolution. To do that we translate the core algorithm of \textsc{FERENGI} \citep{barden2008} into \textsc{python} and we include a more general treatment of the intrinsic luminosity evolution of galaxies as a function of redshift. This treatment ensures that we are using an artificially redshifted sample that closely matches, in brightness, the sample that we have at higher redshifts. The core of the algorithm is shown as a diagram in Figure \ref{fig:dopterian_diagram} and is briefly summarized in the following steps:

\begin{enumerate}
\item We re-scale our image (preserving their total flux) to match the pixel-scale of the high redshift observations while preserving the physical scale of the galaxy.
\item We apply a flux correction to the image that is the combination of two factors: the dimming factor that scales as the inverse of the luminosity distance to the galaxy, and the luminosity evolution factor which account for the average evolution in brightness across different redshifts and is taken from \citet{sobral2013}. Since we are studying rest-frame blue/NUV, we assume light scales with SFR/H$\alpha$.
\item We compute a transformation PSF following the prescription described by \citet{barden2008} applied to the PSFs of the used instruments.
\item We convolve the re-scaled image with the PSF computed in the previous step and then we place the image on an empty region of the target high redshift survey.
\end{enumerate}

Note that we study rest-frame blue/NUV light for a sample of galaxies which are star-forming by selection. So, we expect that at these wavelengths the light will be dominated by young stars and we do not expect great differences across the rest-frame wavelengths that we cover ($2500-5800$\AA).

\begin{figure*}
\centering
\includegraphics[width=\linewidth]{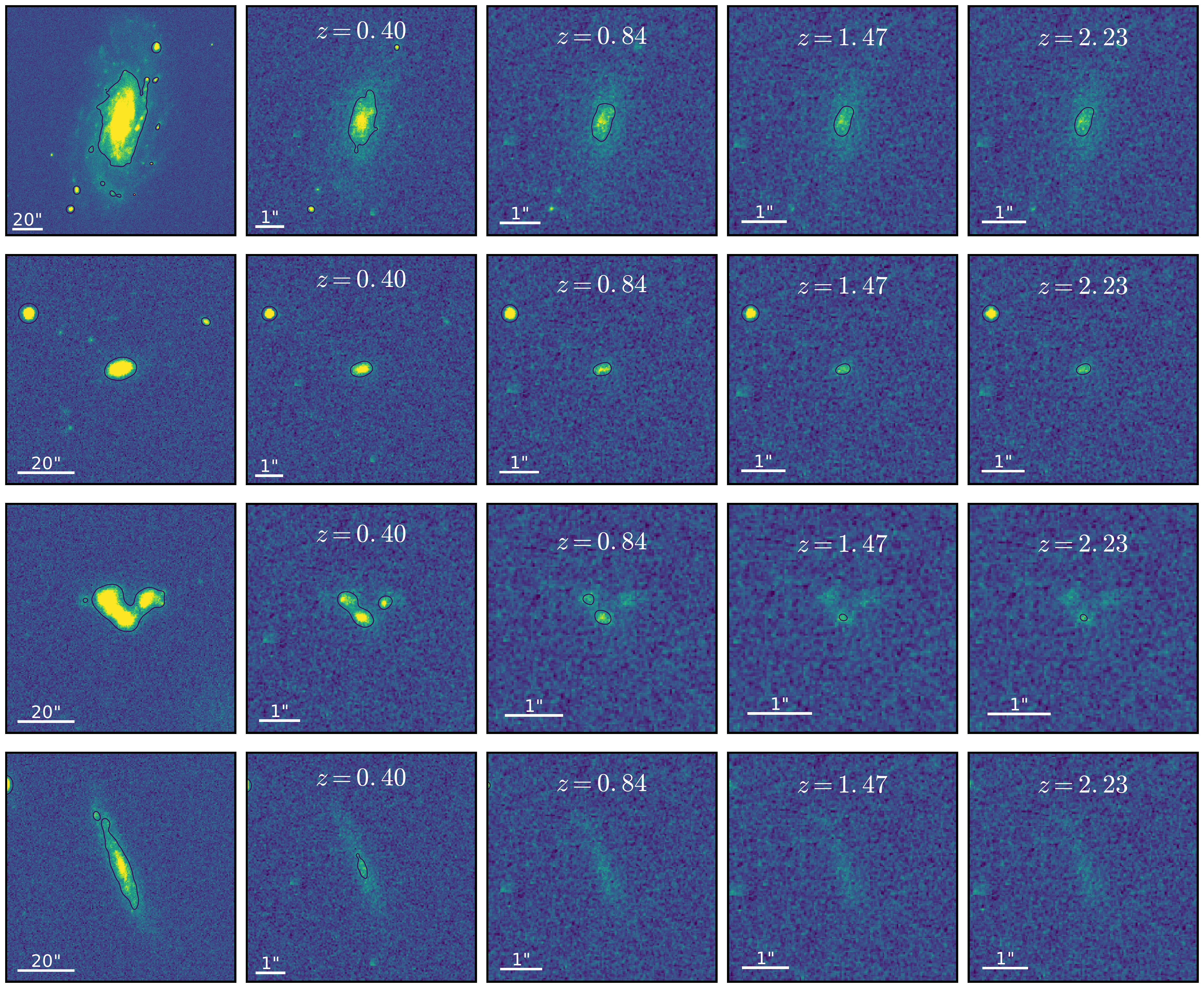}
\caption{Artificial redshifted examples for 4 galaxies from the samples used in this paper. The leftmost panel is the original SDSS $g$-band image. The next 4 panels show the effect of artificial redshifting the galaxy to 4 different redshifts, increasing from left to right. The contours show the 3$\sigma$ level detection in each image. In each image the color map ranges from -3$\sigma$ to $10\sigma$. We observe that the fainter/external regions of the galaxies observed in the local Universe completely disappear as we move towards higher redshifts. All images have the same physical scale. The observed scale is shown for each panel to compare angular sizes at different redshifts.}
\label{fig:example_dopterian}
\end{figure*}

As an example, we show the final result of this algorithm applied in 4 different galaxies redshifted into 4 different redshifts is shown in Figure \ref{fig:example_dopterian}. We show that most of the fine substructures at low redshift are suppressed. The galaxies are still visible, albeit at a low surface brightness level, across all redshifts. 

\section{Parametric Morphological Analysis}\label{section:galfit}

One of the most common ways of characterizing the structure of galaxies is to fit a \citet{sersic1968} profile to the surface brightness distribution of galaxy images \citep[e.g.][]{davies1988,caon1993,andreakis1995,moriondo1998,simard1998,khosroshahi2000,graham2001,mollenhoff2001,trujillo2001,peng2002,blanton2003,trujillo2007,wuyts2011,vanderwel2014}. This profile is characterized by the functional form
\begin{equation*}
I (r) = I_e \exp[-\kappa(r/r_{e})^{1/n}+\kappa],
\label{eq:sersic}
\end{equation*}
where the S\'ersic index $n$ describes the shape of the light profile, $r_e$ is the effective radius of the profile, $I_e$ is the surface brightness at radius $r=r_e$ and $\kappa$ is a parameter coupled to $n$ such that half of the total flux is enclosed within $r_e$. An index of $n=1$ corresponds to a typical exponential disk galaxy, whereas $n=4$ corresponds to the classical de Vaucouleurs profile associated to elliptical galaxies.

To conduct this parametric analysis we make use of \textsc{GALFIT} \citep{peng2002,peng2010}, which is a public algorithm designed to perform a detailed two-dimensional decomposition of galaxies using mathematical models. 

On 2D images, each S\'ersic model has potentially seven free parameters: the position of the center, given by $x_c$ and $y_c$, the total magnitude of the model, $m_{tot}$, the effective radius, $r_e$, the S\'ersic index, $n$, the axis ratio of the ellipse, $b/a$, and the position angle, $\theta_{PA}$, which refers to the angle between the major axis of the ellipse and the vertical axis and it has the sole purpose of rotating the model to match the galaxy's image.

Since \textsc{GALFIT} requires an initial set of parameters from which to start evaluating the model, it is necessary to provide it with realistic guesses. For that reason, we use the source extraction software SExtractor \citep{bertin1996}, which allows one to directly obtain a set of parameters that will serve as input to \textsc{GALFIT} and to find the model that best fits the data faster and with reliable values.

We use $10"\times10"$ cutouts of the HST ACS F814W \citep{scoville2007,koekemoer2007} centered on each HiZELS galaxy. The cutouts are downloaded from the COSMOS HST archive. These images have a typical PSF FWHM of $\sim 0.09$\arcsec, a pixel scale of $0.03\arcsec/\mathrm{pixel}$ and a limiting point-source depth AB(F814W) = 27.2 (5 $\sigma$). For the low-redshift samples, we use images from SDSS DR9 \citep{ahn2012} $g$-band imaging data. These are characterized by a median PSF FWHM of $\sim1.3\arcsec$, a pixel scale of $0.396\arcsec/\mathrm{pixel}$ and a 95\% completeness for point sources at a magnitude of $g=22.2$. For each SDSS image, the cutout size is proportional to the extent of the galaxy we want to fit in order to accommodate a reasonable amount of sky area (at least $50\%$ of the total region). This is done so that \textsc{GALFIT} can simultaneously fit the residual sky emission.

We masked galaxy neighbors by use of the segmentation map output by SExtractor. All sources that fall outside the galaxy of interest are masked. To account for the instrumental PSF, in the case of SDSS data we use interpolated PSF models \citep{lupton2001} reconstructed with \emph{sdss\_psfrec.pro}\footnotemark\footnotetext{\url{http://www.sdss.org/dr12/algorithms/read_psf/}}. For HST F814W images, the PSF was one of the \textsc{TinyTim} \citep{krist1995} models as described by \citet{rhodes2006,rhodes2007}.

Irregular, complex and/or sources detected at low S/N are excluded from the final sample as \textsc{GALFIT} failed to converge on meaningful structural parameters (see Figures \ref{fig:GALFIT_failed_mass} and \ref{fig:GALFIT_failed_lumHa} for a comparison between those and the full sample). These cases amount from 8\% (in the NYU-VAGC sample) up to 40\% (in the $z\sim2.23$ HiZELS sample) of our samples which are distributed in both stellar masses and H$\alpha$ luminosities in the same way as the full sample. Thus, this rate of failure does not introduce any bias against stellar masses or H$\alpha$ luminosities (see Appendix \ref{app:GALFIT_failures}).


\section{Results}\label{section:results}

\subsection{The impact of surface brightness dimming}
\begin{figure*}
\centering
\includegraphics[width=\linewidth]{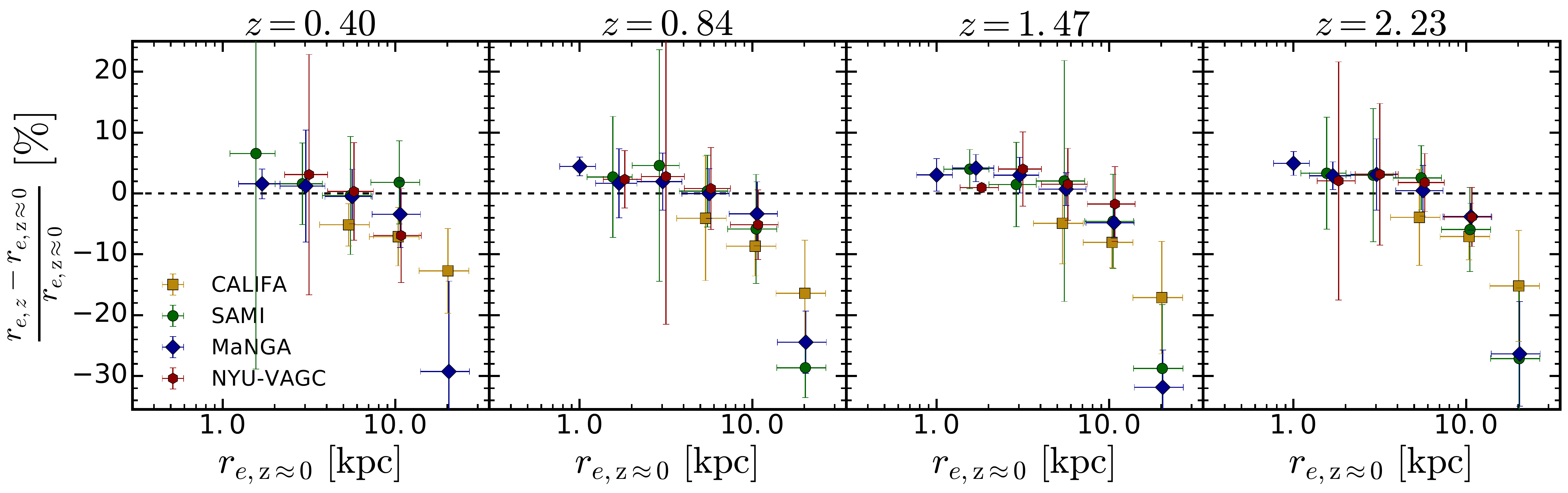}
\caption{The fraction difference between the recovered and input effective radius, $({r_{e,z}}-{r_{e,\mathrm{z\approx0}}})/{r_{e,\mathrm{z\approx0}}}$, for each of the local subsamples and each redshift (one per column). Each symbol represent the median values in bins (with a minimum of $10$ galaxies) of width of the horizontal error bar. The vertical error bar shows the error on the median value. The black dashed line pinpoints a fractional difference of 0, i.e., when the recovered effective radius is the same as input. We find the difference to be very small (<10\%) up to 10 kpc. Above that value, the size of galaxies is more severely underestimated ($\sim10-30\%$). Small horizontal offsets were applied to improve the readability of the plot.}
\label{fig:results_rad_all_redshift_surveys}
\end{figure*}

\begin{figure*}
\centering
\includegraphics[width=\linewidth]{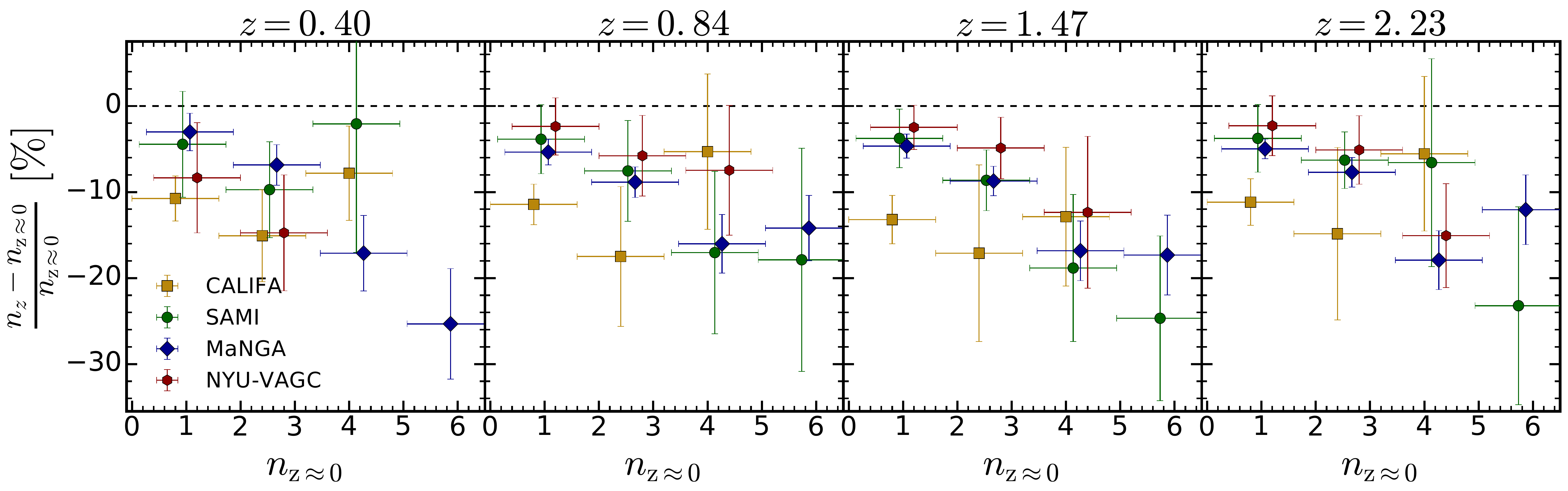}
\caption{The fraction difference between the recovered and input S\'ersic index, $({n_{z}}-{n_{\mathrm{z\approx0}}})/{n_{\mathrm{z\approx0}}}$, for each of the local subsamples and each redshift (one per column). Each symbol represent the median values in bins (with a minimum of $10$ galaxies) of width of the horizontal error bar. The vertical error bar shows the error on the median value. The black dashed line pinpoints a fractional difference of 0, i.e., when the recovered effective radius is the same as input. We find that the value of the S\'ersic index is always underestimated ($\sim2-30\%$) in all redshifts and for all four surveys. Small horizontal offsets were applied to improve the readability of the plot.}
\label{fig:results_sers_all_redshift_surveys}
\end{figure*}

We focus our attention on the impact of surface brightness dimming on two structural parameters that are often used to describe the global morphology of galaxies: the effective radius ($r_e$) and the S\'ersic index ($n$) \citep[see e.g.][]{vanderwel2014,morishita2014,shibuya2015}.

We investigate the ratio of fractional error for $r_e$ for all samples described in Section \ref{ssection:lowz} and at the four redshifts slices observed with HiZELS. We show (in Figure \ref{fig:results_rad_all_redshift_surveys}) that the effective radius is, on average, recovered with success (within an error margin of $\sim10\%$) for galaxies with $r_{e,z\approx0}<10$ kpc. For larger galaxies we reach a saturation value, i.e., galaxies with $r_{e,z\approx0}>10$ kpc are recovered, on average, with $r_e\sim10-20$ kpc (thus the ratio declines for larger radii). This effect is seen on the CALIFA and MaNGA subsamples which are the ones where we have galaxies in this range of sizes. It is also noticeable that for the smaller galaxies we have a higher chance of recovering the value observed at $z\approx0$.

We also investigate the impact on the derived S\'ersic indices. We show in Figure \ref{fig:results_sers_all_redshift_surveys} that we recover the value of the S\'ersic index at a smaller value (by $\sim5-20\%$) than the one that is observed at $z\approx0$. This effect is larger at larger $n_{z\approx0}$ and it shows in all samples at all redshifts. There does not seem to be any systematic offset between different redshifts.

We show in Appendix \ref{app:others} the impact of dimming on the recovered total magnitudes (corrected for the luminosity evolution) and axis ratio of the profiles of individual galaxies. Our results are consistent with those exploring artificial redshifting up to $z\sim1$ \citep{barden2008,mosleh2013} and also for artificial redshifting of massive galaxies from $z\sim0$ to $z\sim2.5$ \citep{weinzirl2011}.

The success of the recovery of the structural parameters for the highest redshift galaxies is directly linked to the counter-balance of the luminosity evolution and the cosmological surface brightness dimming. We have tested a scenario where galaxies undergo no luminosity evolution and the recovery of galaxies (at the largest radius) is severely affected. This would imply that we would not be able to observe large galaxies at higher redshifts and thus, our distributions would be skewed towards smaller sizes potentially mimicking an evolution scenario.

That impact is explored in Table \ref{tab:median_fraction} where we compare the median values of the distributions in $r_e$ and $n$ at each redshift compared to the observed median value at $z\approx0$. As somewhat expected from the comparison of individual galaxies, we observe no systematic trend induced by the cosmological dimming on the recovered median values. For $r_e$ we see that we slightly underestimate sizes for galaxies in the CALIFA sample ($\sim 90\%$ of the original). For the other samples, we get the same median value within $\lesssim10\%$ at all redshifts. As for the value of $n$, we systematically underestimate the median value at a level of $11-27\%$ of the value observed at $z\approx0$ with all artificially redshifted samples. This value does not seem to change as a function of redshift.

We discuss the dependence of the fractional error of sizes and S\'ersic indexes on the input magnitude in Appendix \ref{app:others}. The reader is also referred to Appendix \ref{app:others}  for a discussion on the possible explanations for a systematic overestimation of sizes due to dimming.

\begin{table*}
\centering
\begin{tabular}{ |c|c|c|c|c| }
\hline
  & \multicolumn{4}{c}{$\frac{r_{e,z}}{r_{e,\mathrm{z\approx0}}} ~|~ \frac{n_{z}}{n_{\mathrm{z\approx0}}}$} \\
\hline
  & $z=0.40$ & $z=0.84$ & $z=1.47$ & $z=2.23$ \\
 \hline
CALIFA & 0.885 | 0.805 & 0.918 | 0.828 & 0.863 | 0.829 & 0.918 | 0.839  \\
SAMI & 1.006 | 0.869 & 1.166 | 0.763 & 1.096 | 0.812 & 1.052 | 0.789  \\
MaNGA & 1.031 | 0.856 & 1.037 | 0.848 & 1.039 | 0.851 & 1.037 | 0.832  \\
NYU-VAGC & 1.107 | 0.729 & 1.011 | 0.841 & 0.989 | 0.832 & 1.001 | 0.887  \\
\hline
Average & 1.007 | 0.815& 1.033 | 0.820& 0.997 | 0.831& 1.002 | 0.837\\
\hline
\end{tabular}
\caption{Ratio of the median values, $r_e$ and $n$, for each simulated redshift to the observed values at z$\approx$0. This dimensionless value quantifies the over/under estimation of each parameter as a function of the simulated redshift. The average values derived from the 4 samples are used for correction.}
\label{tab:median_fraction}
\end{table*}

\subsection{The structure and sizes of galaxies}
\begin{figure}
\centering
\includegraphics[width=\linewidth]{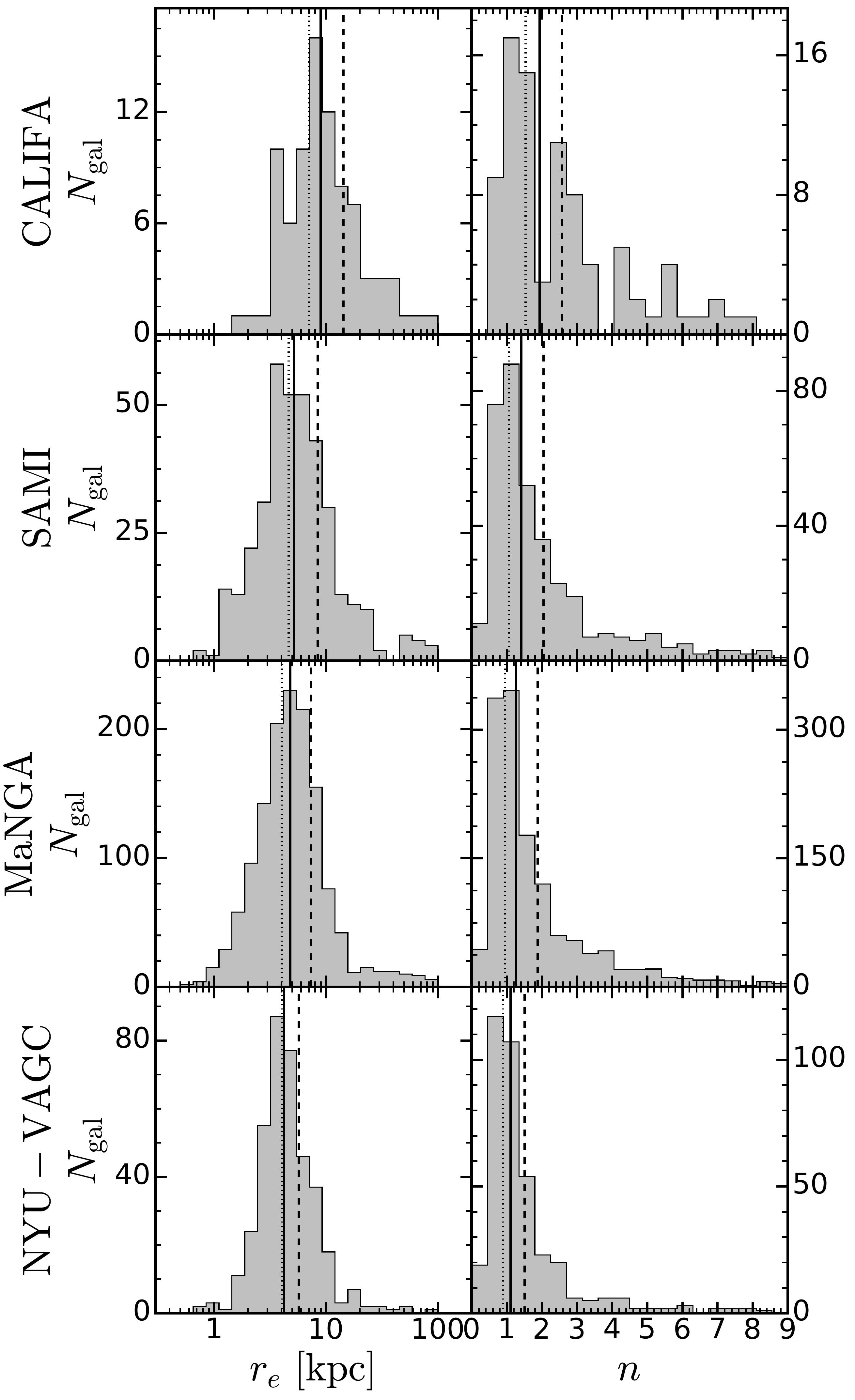}
\caption{Distributions of sizes ($r_e$) and S\'ersic indices ($n$) of the galaxies observed at $z\approx0$ for the $4$ local samples defined in Section \ref{section:sample}. The vertical solid, dashed an dotted lines indicate the median, mean and modal values, respectively.}
\label{fig:results_galfit_local}
\end{figure}

\begin{figure}
\centering
\includegraphics[width=\linewidth]{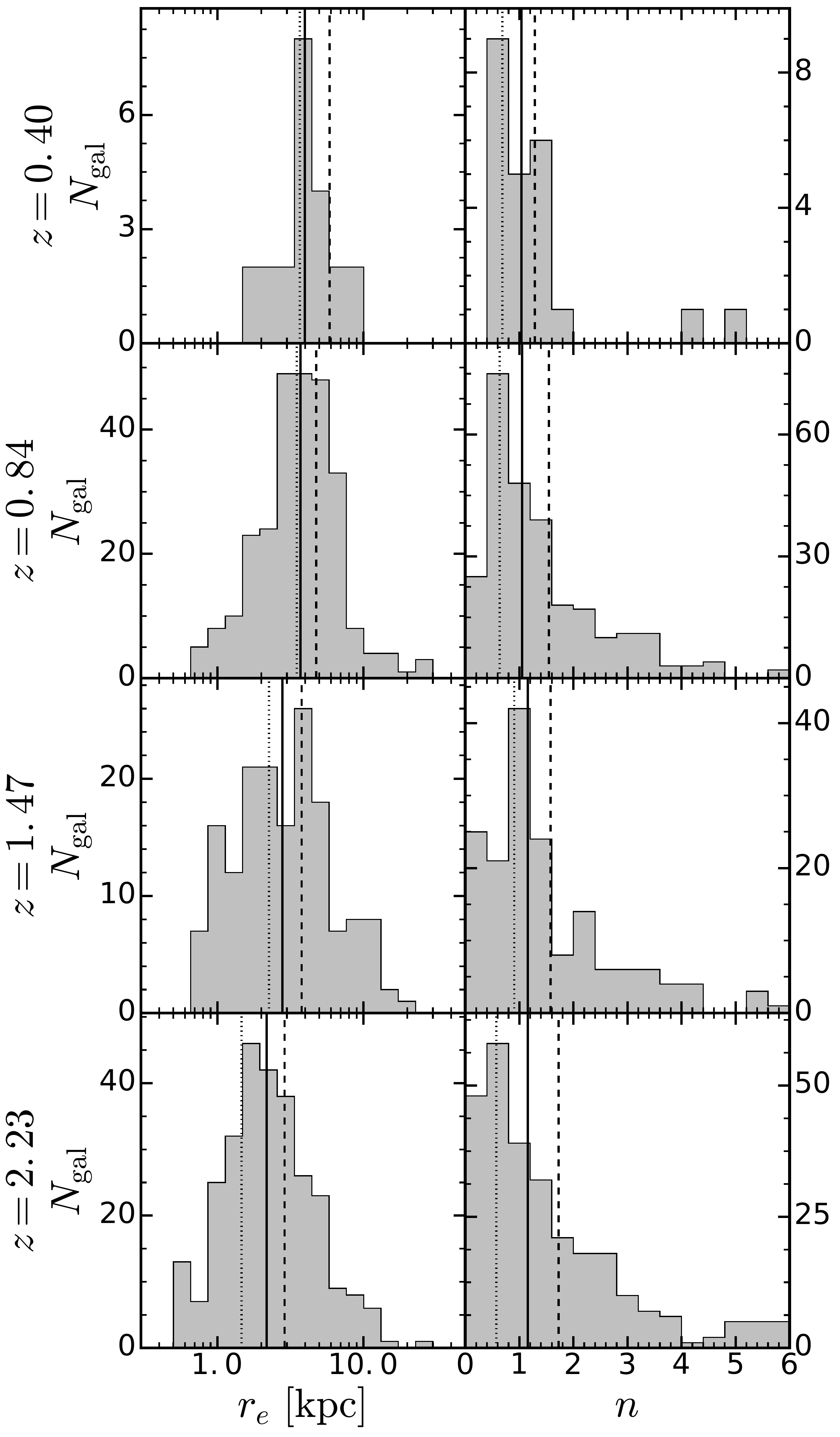}
\caption{Distributions of sizes ($r_e$) and S\'ersic indices ($n$) of the galaxies observed at high redshift from HiZELS. The vertical solid, dashed an dotted lines indicate the median, mean and modal values, respectively. The distributions are based on the observed values without any correction.}
\label{fig:results_galfit_hizels}
\end{figure}

\begin{figure}
\centering
\includegraphics[width=\linewidth]{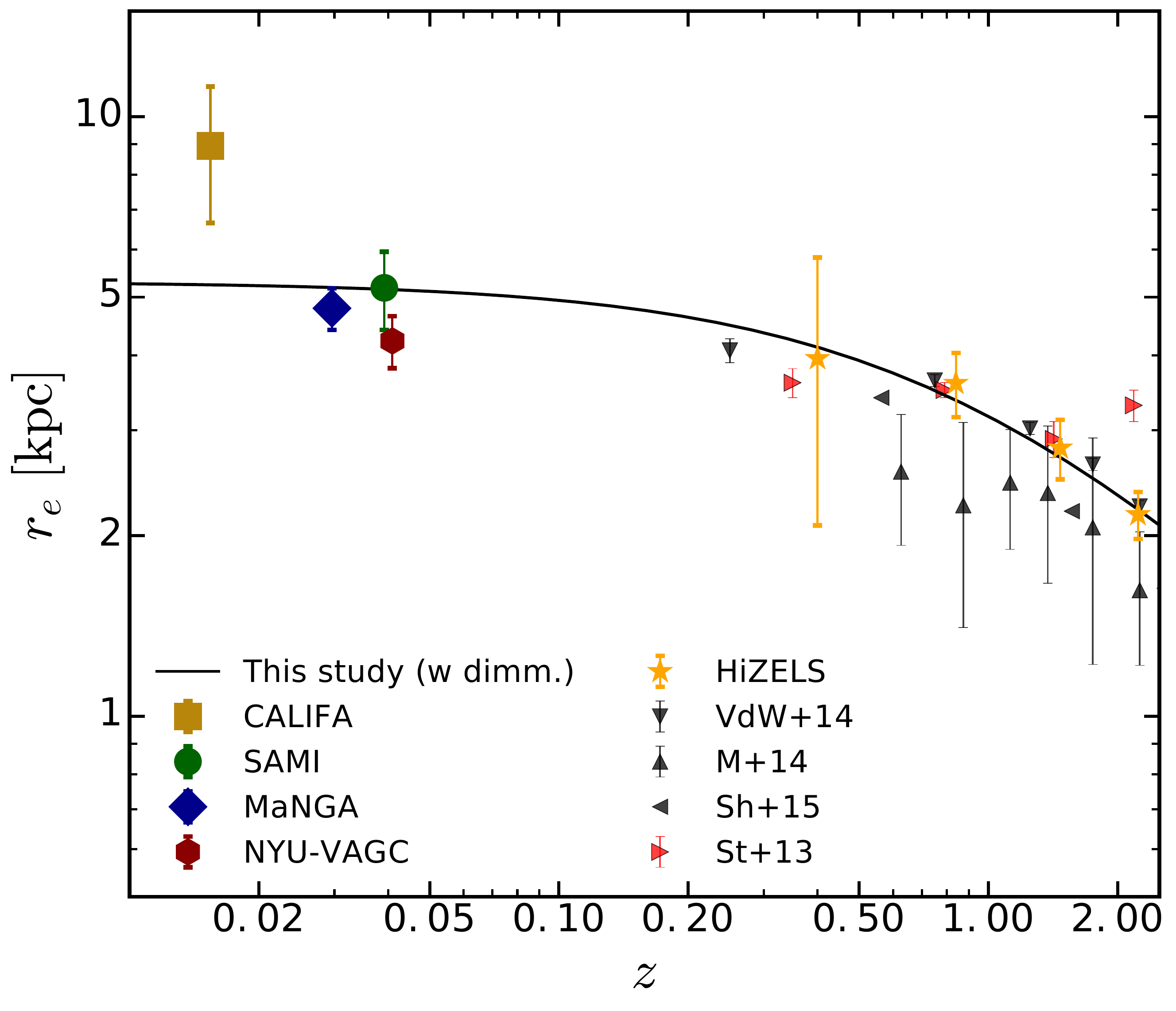}
\includegraphics[width=\linewidth]{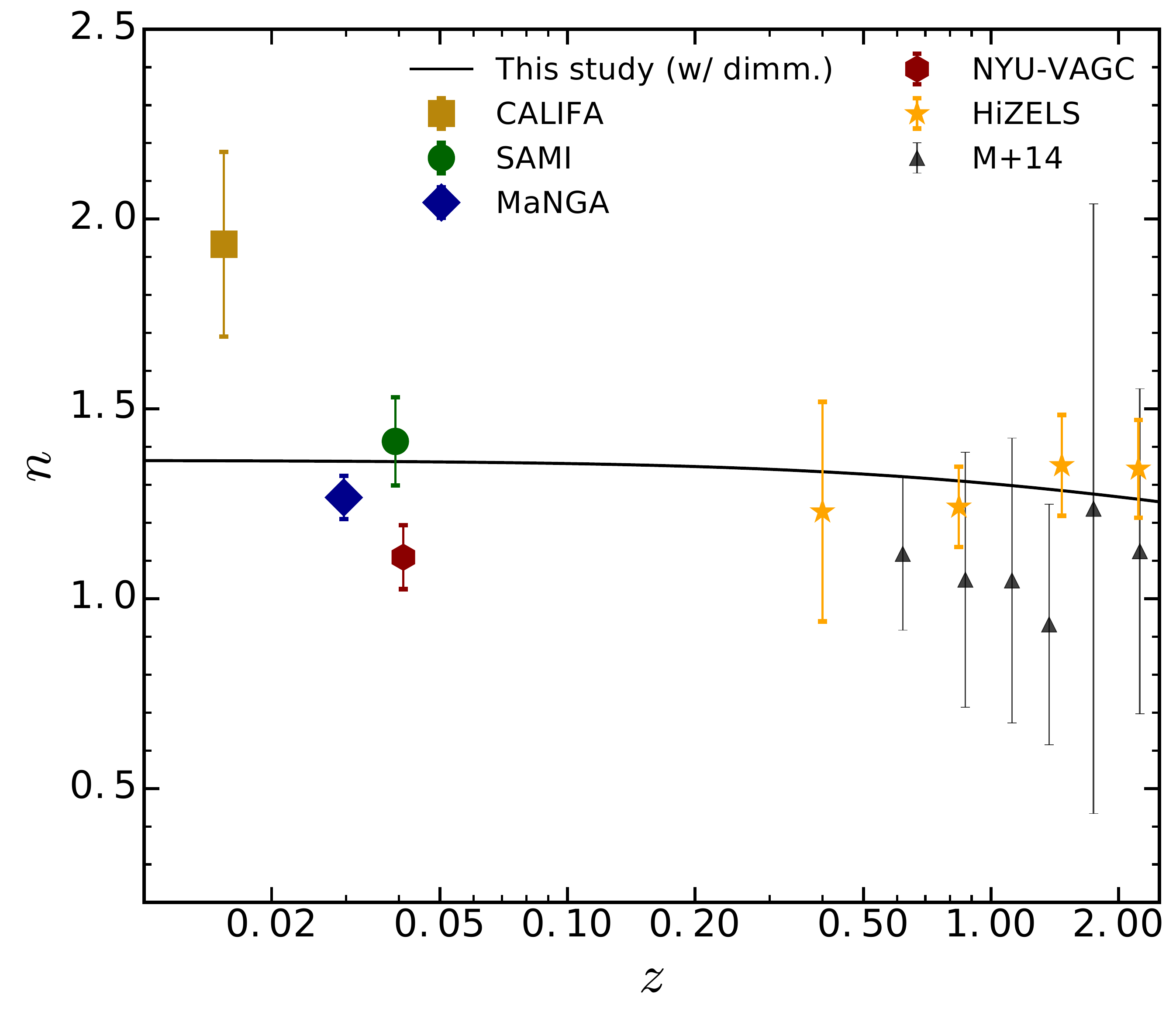}
\caption{Structural evolution of SFGs as parameterized by S\'ersic profiles. Top: the evolution of galaxies' median (observed) effective radius as a function of the median redshift. Bottom: same as top but for the galaxies' median S\'ersic index. The colored points show the values (after dimming correction for HiZELS) derived from this study. The triangles show reported values from \citet{stott2013b,vanderwel2014,morishita2014,shibuya2015}. The points from \citet{stott2013b} have been horizontally offset for viewing purposes. The solid line shows the fit to equation \ref{eq:zevo} after dimming correction.}
\label{fig:results_average_galfit}
\end{figure}

\begin{table*}
\centering
\begin{tabular}{ |c|c|c|c|c|c|c| }
\hline
          & &$r_e [\mathrm{kpc}]$& & & $n$& \\
\cmidrule(lr){2-4} \cmidrule(lr){5-7}
Sample & median & mean & mode & median & mean & mode \\
\hline
CALIFA & $8.94^{+11.13}_{-4.60}$ & $14.30 \pm 1.83$ & 7.08 & $1.93^{+2.34}_{-0.85}$ & $2.57 \pm 0.19$ & 1.53 \\
SAMI & $5.18^{+5.54}_{-2.51}$ & $8.40 \pm 0.62$ & 4.65 & $1.41^{+1.89}_{-0.67}$ & $2.05 \pm 0.09$ & 1.06 \\
MaNGA & $4.79^{+3.94}_{-2.24}$ & $7.34 \pm 0.30$ & 4.03 & $1.27^{+1.83}_{-0.56}$ & $1.88 \pm 0.05$ & 0.95 \\
NYU-VAGC & $4.23^{+3.30}_{-1.50}$ & $5.72 \pm 0.34$ & 4.05 & $1.11^{+1.07}_{-0.45}$ & $1.51 \pm 0.07$ & 0.89 \\
$z=0.40$  & $3.98^{+2.34}_{-1.30}$ & $5.88 \pm 1.49$ & 3.67 & $1.04^{+0.49}_{-0.47}$ & $1.29 \pm 0.23$ & 0.69 \\
$z=0.84$  & $3.72^{+2.45}_{-1.87}$ & $4.77 \pm 0.35$ & 3.50 & $1.05^{+1.49}_{-0.55}$ & $1.55 \pm 0.08$ & 0.64 \\
$z=1.47$  & $2.79^{+2.89}_{-1.55}$ & $3.78 \pm 0.25$ & 2.26 & $1.16^{+1.62}_{-0.72}$ & $1.58 \pm 0.11$ & 0.91 \\
$z=2.23$  & $2.17^{+2.35}_{-1.11}$ & $2.89 \pm 0.16$ & 1.47 & $1.15^{+1.92}_{-0.76}$ & $1.73 \pm 0.10$ & 0.58 \\
\hline
\end{tabular}
\caption{Median, mean and modal values for the effective radius ($r_e$) and S\'ersic index ($n$) for all samples described in Section \ref{section:sample}. The errors on the median denote the 16th (lower bound) and 84th (upper bound) percentiles of the variables distribution.}
\label{tab:median_values}
\end{table*}

\begin{table}
\centering
\begin{tabular}{ |c|c|c|c| }
\hline
Variable & Correction &All samples  & High-$z$ only\\
\hline
\multirow{2}{*}{$\alpha_{r_e}$} &w/o dimm. & $-0.74 \pm 0.20$ &  $-0.71 \pm 0.11$ \\
   &w/ dimm.& $-0.75 \pm 0.20$ & $-0.70 \pm 0.09$ \\
\hline
\multirow{2}{*}{$\alpha_{n}$} &w/o dimm. & $-0.25 \pm 0.21$ &  $0.15 \pm 0.04$ \\
  &w/ dimm. & $-0.07 \pm 0.19$ & $0.12 \pm 0.04$ \\
\hline
\end{tabular}
\caption{Median values of $\alpha_{X}$ after fitting through all samples and for fitting only the HiZELS high-$z$ sample.}
\label{tab:slope_evolution}
\end{table}
 
We plot the histograms of sizes ($r_e$) and S\'ersic indices ($n$) in figures \ref{fig:results_galfit_local} and \ref{fig:results_galfit_hizels} and we summarize the median values in Table \ref{tab:median_values}. For the local samples we see that galaxies have, on average,  $r_e\sim9$ kpc for the CALIFA sample and $r_e\sim4-5$ kpc for the other three samples. We want to stress that the CALIFA sample is diameter selected to match the diameter of the IFU field of view and that is what is driving the larger sizes with respect to the other samples, which have lesser to no constraints on galaxy size in their selection. As for the values of $n$, we see that the distributions are similar, but peaking at increasing values of $n$ for NYU-VAGC, MaNGA, SAMI and CALIFA, respectively. Again, the CALIFA sample shows an increased fraction of galaxies with high value of $n$, which is likely due to the morphological constraint to cover the entire Hubble diagram.

As we examine the high redshift galaxies from HiZELS (Figure \ref{fig:results_galfit_hizels}) we see that galaxies have increasingly smaller sizes as we move towards higher redshifts with median values ranging from $r_e\sim2$ kpc at $z=2.23$ up to $r_e\sim4$ kpc at $z=0.40$. As for the values of the S\'ersic indices, we see that it is rather stable at these redshifts with median values around $n\sim1.1$, close to exponential disk profiles characteristic of SFGs in the local Universe. We do not observe an enhancement of the fraction of higher $n$ galaxies at any redshift probed with HiZELS. We note that this is not driven by observational biases.

There are some cases of SFGs with large sizes and/or S\'ersic indices seen in Figure \ref{fig:results_galfit_hizels}. However, these generally come from poorly constrained fits, which involve large errors on both $r_e$ and $n$.

\subsection{The evolution of the structure and sizes of galaxies}

After obtaining the structural parameters of our samples we can see how they compare in terms of their evolution. In Figure \ref{fig:results_average_galfit} we show how the measured sizes and S\'ersic indices depend on redshift. We fit the equation
\begin{equation}
X \propto (1+z)^{\alpha_{X}}
\label{eq:zevo}
\end{equation}
for $X=r_e,n$ to quantify the rate of change of these quantities across cosmic time. We compute these quantities by considering, or not, the impact of dimming in terms of the overestimate/underestimate average ratios shown in Table \ref{tab:median_fraction}. Theses results are summarized on Table \ref{tab:slope_evolution}.

From an overall perspective on the H$\alpha$-selected sample the scenario of size growth is compatible with mild evolution ($\alpha_{r_e}\approx-0.7\pm0.11$) between redshifts $2.2$ and $0.4$. We see in Table \ref{tab:slope_evolution} that considering the entire redshift range (by including the local samples in the fit) increases slightly the value of the slope to $\alpha_{r_e}\approx-0.75\pm0.20$, which is fully consistent with the result obtained using only the higher redshift samples from HiZELS. We can also see that including the dimming correction has virtually no impact on the derived value of $\alpha_{r_e}$, as expected from what we see in Table \ref{tab:median_fraction}. These slopes deviate more than 3$\sigma$ from the no evolution scenario ($\alpha=0$) and thus it is very unlikely that galaxies experience no growth across this period.

We also investigate any potential evolution of the median value of the S\'ersic index of galaxies where we get slightly different solutions if we look only at the higher redshift samples (decreasing S\'ersic index with time, $\alpha_n\approx0.15\pm0.04$) or include all the samples in the fit (increasing S\'ersic index with time, $\alpha_n\approx-0.25\pm0.21$). However, we note that the slopes are very shallow and point to almost no evolution. These two slightly contrasting scenarios are appeased when we include the impact of dimming in the derived value of $n$ derived in Table \ref{tab:median_fraction}. With that correction we obtain $\alpha_n\approx-0.07\pm0.19$ for all samples and $\alpha_n\approx0.12\pm0.04$ when only considering the HiZELS samples. Considering the full range in redshifts, our derived evolution encompasses the value $\alpha=0$ within the $1\sigma$ error. This supports a scenario where SFGs maintain their surface brightness distribution close to pure exponential disks at all times.

\subsection{The size-mass relation across cosmic time}

\begin{figure}
\centering
\includegraphics[height=0.800\textheight]{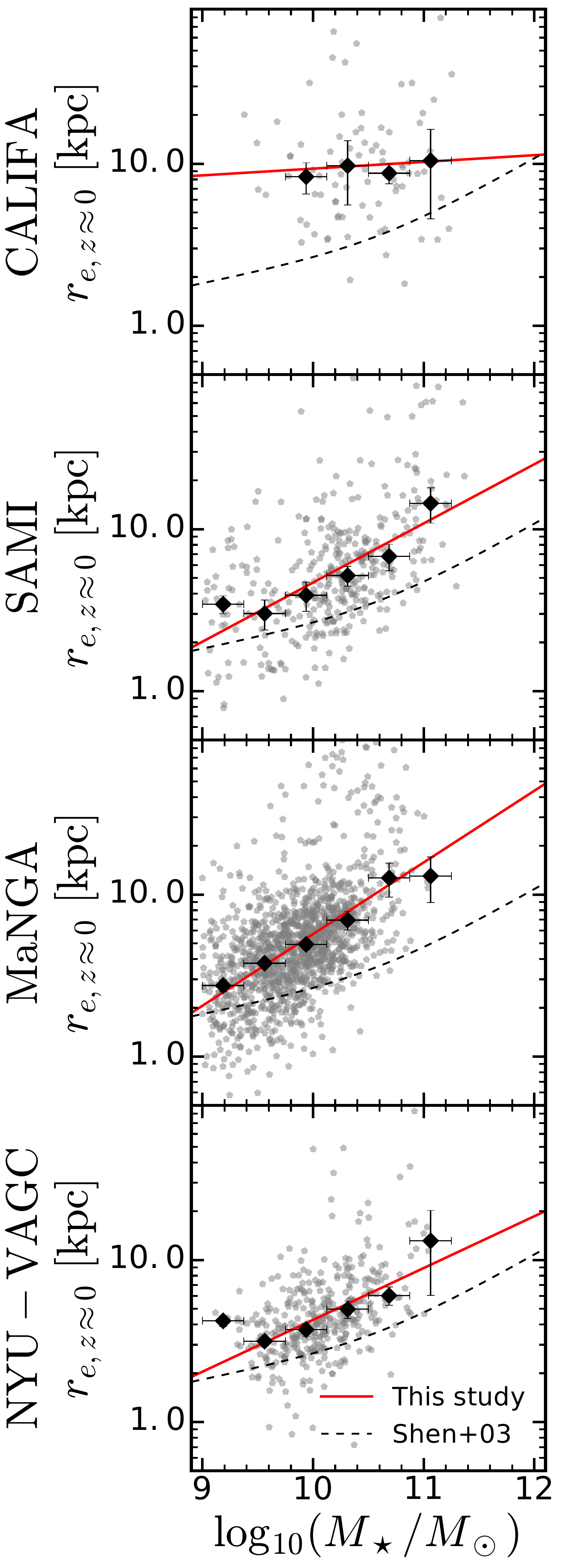}
\caption{The stellar mass-size relation for the local Universe samples described in Section \ref{section:sample}. Each gray pentagon represents an individual galaxy, the large black diamonds are the median values in a stellar mass bin of width indicated by the horizontal error bar. The vertical error bar denotes the error on the median. The red solid line shows the best fit of equation \ref{eq:mass-size} to the data (see also Table \ref{tab:slope_size_mass}). We include the fits from \citet{Shen2003} as a solid dashed line. We note that the reason for the differences between our fits and those by \citet{Shen2003} is mainly due to the use of different size estimator. We use the major axis effective radius and they use the ciruclarized effective radius. When using the latter in our data we find a better agreement. See Sect. \ref{section:discussion} for more details.}
\label{fig:results_mass_size_local}
\end{figure}

\begin{figure}
\centering
\includegraphics[height=0.800\textheight]{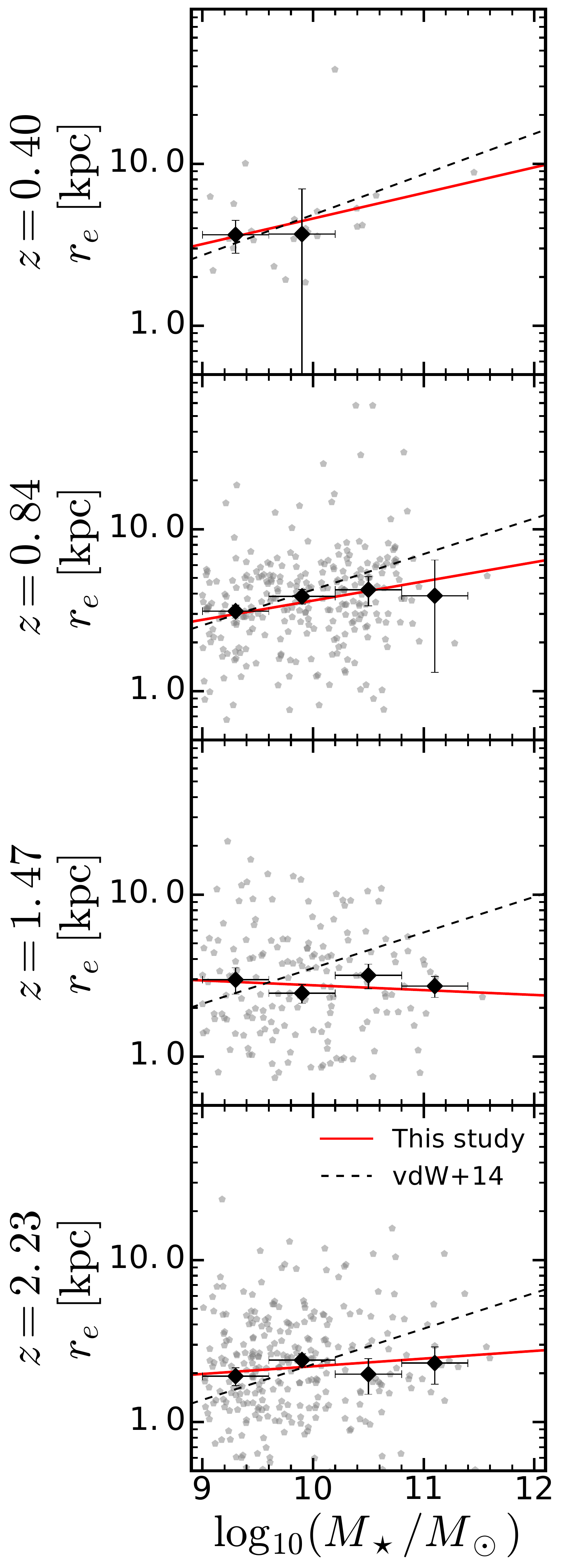}
\caption{The stellar mass-size relation for 4 redshifts probed by HiZELS. Each gray pentagon represents an individual galaxy, the large black diamonds are the median values in a stellar mass bin of width indicated by the horizontal error bar. The vertical error bar denotes the error on the median. The red solid line shows the best fit of equation \ref{eq:mass-size} to the data (see also Table \ref{tab:slope_size_mass}). We include the fits from \citet{vanderwel2014} as a solid dashed line (from top to bottom, the redshift bins from that paper are: $0.0<z<0.5$, $0.5<z<1.0$, $1.0<z<1.5$, $2<z<2.5$).}
\label{fig:results_mass_size_hizels}
\end{figure}

\begin{figure}
\centering
\includegraphics[width=\linewidth]{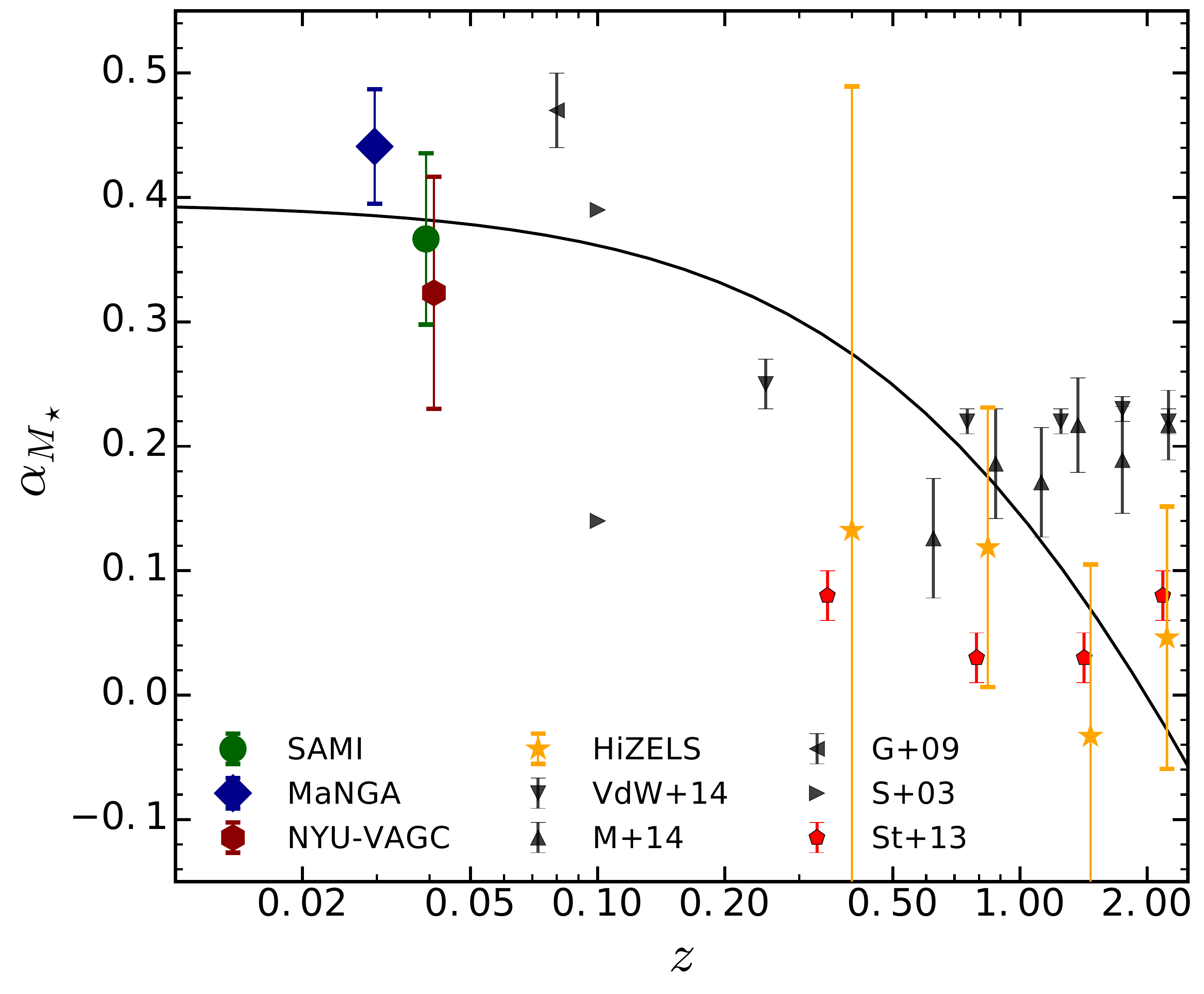}
\caption{The value of the slope of the stellar mass-size relation as a function of the median redshift of each sample. Larger full colored symbols are for the samples studied in this paper. The semi-transparent triangles are from other reports from the literature: \citet{Shen2003,Guo2009,vanderwel2014,morishita2014}. \citet[]{Shen2003} symbols correspond to the low ($0.14$ at $<10^{10.6}M_\odot$) and high ($0.39$ at $>10^{10.6}M_\odot$) stellar mass end. Values from \citet{stott2013b} are shown as red pentagons and a horizontally shift was applied for viewing purposes}. The solid line shows the best fit through our data points of $\alpha_{M_\star} = A\log_{10}(1+z)+B$, with $A=0.8\pm0.2$ and $B=0.40\pm0.05$.
\label{fig:results_mass_size_slope_evolution}
\end{figure}

The other aspect to retain from the morphological analysis is the relation of the structural parameters with the stellar mass, which is displayed in figures \ref{fig:results_mass_size_local} and \ref{fig:results_mass_size_hizels}. There, we show the median values in bins of stellar mass and their dispersion on the stellar mass-size relation alongside with the individual measurements for all galaxies. In order to parameterize this relation we use the functional form
\begin{equation}
r_e \propto M_\star^{\alpha_{M_\star}}
\label{eq:mass-size}
\end{equation}
which is over plotted as a red solid line in those figures. The slopes of this relation are summarized in Table \ref{tab:slope_size_mass}.

To fit the stellar mass-size relation we perform a Monte Carlo Markov Chain simulation \citep[using the \emph{emcee} package, ][]{emcee2013} where all individual galaxies are considered for the fit. The first guess is computed from a simple fit to the binned points. We use a total of 50 chains with initial guesses randomly deviated from the simple fit. Each chain then runs for 2000 steps and we exclude the first 500 steps for each chain to erase the influence of the initial first guesses. The reported errors on the slope are the width of the posterior distribution.

\begin{table}
\centering
\begin{tabular}{ |c|c| }
\hline
Sample & $\alpha_{M_\star}$\\
\hline
CALIFA & $0.05 \pm 0.17$ \\
SAMI & $0.37 \pm 0.07$ \\
MaNGA & $0.44 \pm 0.05$ \\
NYU-VAGC & $0.32 \pm 0.09$ \\
$z=0.40$ & $0.15 \pm 0.36$ \\
$z=0.84$ & $0.11 \pm 0.11$ \\
$z=1.47$ & $-0.03 \pm 0.14$ \\
$z=2.23$ & $0.05 \pm 0.11$ \\
\hline
\end{tabular}
\caption{Slope of the stellar mass-size relation for all the samples described in Section \ref{section:sample}.}
\label{tab:slope_size_mass}
\end{table}

We show that for the local samples, apart from CALIFA, there is a more evident correlation of sizes with stellar masses as that measured from the HiZELS samples. Again, for the CALIFA sample, the marked difference against the other local Universe samples is related to the size constraint imposed for the parent selection of CALIFA. This results in a flatter stellar mass-size relation as galaxies were chosen to have similar sizes within the sample and stresses the importance of sample selection whenever we study a galaxy population. Due to this selection effect, we disregard the CALIFA sample when establishing comparisons between local and high-redshift morphologies. Nonetheless, despite the apparent inversion of the relation for $z=1.47$, we observe that on average the most massive galaxies are larger than their lower mass counterparts. 

We compare our best fit relations with those found in the literature for SFGs and find that in the local Universe the CALIFA sample is the one that deviates the most from what is found in SDSS by \citet{Shen2003} and \citet{Guo2009}. This deviation from CALIFA is expected since it is a size-selected sample and thus a flatter correlation with stellar mass is unsurprising. As we move towards higher redshifts, we observe that the HiZELS sample changes its slope to shallower values. This is in contrast with the apparent constancy in the stellar mass-size relation slopes found by \citet{vanderwel2014} and \citet{morishita2014}. We note that their selection of SFGs is based on a color-color diagram, and not on H$\alpha$. On top of that, we measure our relation on bluer wavelengths than those studies which are based on rest-frame optical data on smaller areas of the sky.

We go further in the investigation of the evolution of this stellar mass-size relation across cosmic time and plot the values of the slope as a function of redshift in Figure \ref{fig:results_mass_size_slope_evolution}. Again, if one excludes CALIFA sample (due to its selection bias), the relation is becoming steeper as we move towards the local Universe. We note also that the steepening appears to occur mostly between $z=0.$4 and $z\approx0$ as when we look only at the HiZELS samples the rate of change on the value of $\alpha_{M_\star}$ is smaller. We have attempted to improve the robustness of our fits by relaxing the H$\alpha$ luminosity cut and down to 0.01$L^*_\mathrm{H\alpha}$ for $z=0.4$. We have 95 galaxies in this scenario and derive a slope of $\alpha_M = 0.17\pm0.26$, consistent with what we have but with a slightly smaller error.

By using the fits derived in equation \ref{eq:mass-size} it is possible to derive the effective radius at a given stellar mass for all the samples we are studying in this paper. We have computed the sizes of galaxies at $\log_{10} M_\star/M_\odot = 10,10.5$ and find that they follow a similar trend as displayed in Figure \ref{fig:results_average_galfit} and Table \ref{tab:median_values}. The evolution one would derive from this quantity is similar to that of using the median sizes for the galaxy population. Additionally one can see that the more massive galaxies tend to grow faster with cosmic time as the differences between consecutive redshift slices is larger at $\log_{10} M_\star/M_\odot =10.5$ than at $\log_{10} M_\star/M_\odot = 10$.

\subsection{Impact of cosmological dimming on the stellar-mass-size relation}

We attempt to quantify the impact that dimming may have on the derived stellar mass-size relations by exploring the dependence of the fractional error on the effective radius on the stellar mass of the input galaxy. In Figure \ref{fig:results_radius_vs_mass_all_redshift_surveys} we see that there is no strong correlation between the two quantities. However, we do find that galaxies at the highest masses ($\log_{10}(M_\star/M_\odot)\gtrsim10.5$) are more likely to have their sizes underestimated at a level of $\lesssim 20 \%$. We do not find any dependence on the effect of dimming with simulated redshift slice. These two results imply that it is unlikely that the shallow slopes that we observe for our stellar mass-size relations are caused by the cosmological dimming.

Thus, we believe that the differences we observe are due to our selection based on H$\alpha$ and to the weaker correlation between rest-frame UV light and stellar mass which is measured mostly from rest-frame NIR.


\section{Discussion}\label{section:discussion}

In order to infer the true evolution of galaxies we need to account for any observational bias that may affect our observed results. Regarding morphology, and its evolution with redshift, the strongest bias could come from cosmological surface brightness dimming. The impact of this dimming was already extensively explored out to $z\sim1-2.5$ by \citet{barden2008} \citep[see also][]{trujillo2007,franx2008,weinzirl2011,mosleh2013}, where they find that to have little impact on \textsc{GALFIT} based measurements in this redshift range. Nevertheless due to the strong dependence of dimming on redshift and the luminosity evolution of galaxies with redshift, any evolutionary trend must be taken with care. For instance, \citet{ichikawa2012} found a small evolution on galaxy sizes in optical rest-frame, but consistent with the expected effect from cosmological dimming. We show in Section \ref{section:results} that indeed the cosmological dimming has a small impact on the derived sizes and it is more important on the derived S\'ersic indices.

As for the local samples we find galaxies that are 2-3 times larger, depending on the local sample we consider, when compared to other studies based on late-type galaxies in SDSS. For example, \citet{Shen2003} find values of the half-light Petrosian radius $r_{50}\sim2.5$ ($3.2$) kpc at stellar masses of $\log_{10}(M_\star/M_\odot)=10$ ($10.5$). \citet{Guo2009} study central galaxies in SDSS, and find that late-type galaxies have $\sqrt{b/a}r_{e}\sim2.45$ ($4.78$) kpc at stellar masses of $\log_{10}(M_\star/M_\odot)=10$ ($10.5$). 

The differences that we find in the typical sizes of galaxies are a reflection of differences in the overall stellar mass-size relation. \citet{Shen2003} fitted a double power law and finds slopes of $\alpha_{M_\star}=0.14$ for $\log_{10}(M_\star/M_\odot)<10.6$ and $\alpha_{M_\star}=0.39$ for $\log_{10}(M_\star/M_\odot)>10.6$. Note that we use the major axis effective radius in all our plots to be comparable to those reported by the literature at high-redshift, namely the values from \citet{stott2013b}. However, the values reported by \citet{Shen2003} refer to the circularized effective radius. When using this size estimate we get a much better agreement between our results and those reported by  \citet{Shen2003}, especially on the control NYU-VAGC sample. We do, however, still find a bias that IFU samples have slightly larger galaxies at high masses. This small bias may be perceived as an attempt to maximize the covering factor of IFU instruments. We computed the absolute difference between between our best fit model and the quoted best fit by \citet{Shen2003} in the stellar mass interval $9<\log(M_\star/M_\odot)<11.5$ we find that difference to be a factor of 2-3 smaller when using the circularized effective radius.

\citet{Guo2009} finds a steeper slope of the stellar mass-size relation with $\alpha_{M_\star}=0.47\pm0.03$, which is slightly higher but statistically compatible with our observed slopes for the local samples. We note that our SFGs are lying in between the slopes that are found but have larger sizes at the same stellar masses. However, we stress that the selection of late-type galaxies in SDSS is different than what we have applied (see Section \ref{section:sample}). They are based in either visual inspection \citep{Guo2009} and structural separation \citep[using the concentration $c<2.86$ and $n<2.5$,][]{Shen2003}. We also exclude low mass galaxies from the fit ($\log(M_\star/M_\odot)<9$), which influence the stellar mass-size relation by \citet{Shen2003}. We are also missing very massive galaxies ($\log(M_\star/M_\odot)\gtrsim11$), which could influence our slopes to shallower values if they populate a lower size than what is predicted from our fits. We note however that we agree with the results from the literature at a $\sim2 \sigma$ level and that our shallower slopes could be driven by our smaller sample size or that we are measuring sizes in the blue-NUV rest-frame region.

As for the size evolution our derived trends are consistent with the slopes found by \citet{vanderwel2014}, $\alpha_{r_e}=-0.75$ and slightly steeper than what was found by \citet{morishita2014}, $\alpha_{r_e}=-0.57$. Both these studies target SFGs selected from the $UVJ$ diagram \citep{williams2009} at redshifts $z<3.0$. They also focus on the stellar mass-size relation and find a consistent slope of  $\alpha_{M_\star}=0.22$ \citep{vanderwel2014} and $\alpha_{M_\star}\sim0.2$ \citep{morishita2014} at all redshifts. This slopes are within the errors of the slope found at $z<0.84$, but we find consistently shallower slopes at higher redshifts and a possible hint of evolution with increasingly shallow slopes as we move to higher redshifts.

These shallow slopes are more consistent with those derived by \citet{stott2013b}, using the same HiZELS sample but measuring sizes in rest-frame ground-based optical images. They do not find any evolution at the same redshifts and find remarkably constant sizes with $r_e(M_\star=10^{10}M_\odot)\sim3.6$ kpc in the same redshifts we probe here. The small change in sizes found by \citet{stott2013b} may seem contradictory to the evolution we find which is mostly anchored on the larger difference in sizes observed at $z=2.23$. We believe that the different findings may be caused by a different sample selection but more importantly by the different resolution and rest-frame bands that we use. \citet{stott2013b} use of $K$ band imaging data (covers the region $6800-15700$\AA\ rest-frame) and can be less prone to a morphological $k$-correction than the use of F814W which covers the region $2500-5800$\AA\ rest-frame). This is  especially true for galaxies with a strong $D_n4000$\AA\ break and for galaxies with spatially disparate young and old stellar populations. However, for strong SFGs, the break is expected to be small \citep[e.g. Figure 2 of][and references therein]{cheng2015}. Additionally, it has been shown that for galaxies with dominant star-forming population the morphological $k$-correction is small \citep[e.g.][]{conselice2000c,windhorst2002,taylor-mager2007}. Nonetheless, it is possible that we are seeing a different path of evolution for the young star-forming regions when compared to the older underlying stellar population. This points to a scenario where star formation happens from inside-out and we are witnessing the star formation activity extending to the outer regions of the galactic disk \citep[see e.g.][]{hagen2016}.

\citet{vanderwel2014} finds that late-type galaxies are smaller at longer wavelengths and that this trend is stronger for the most massive galaxies. This trend is weaker as we move towards higher redshifts. When applied to the HiZELS sample, where the provided corrections can be applied, these two effects combined could contribute to shallower stellar mass-size relation slopes as the most-massive, lower redshift galaxies would be the ones with the largest decrease in size. However, we already find shallower slopes than those reported by other studies and these effects would only contribute to a more consistent, no-evolution scenario, of the slope of the stellar mass-size relation which would then be consistent with what is found by \citet{stott2013b}.

We can also compare our results to other typical selections of high-redshift galaxies, namely those based on the Lyman Break Galaxies \citep[LBGs,][]{steidel1996}. For example, \citep{ferguson2004,mosleh2011,shibuya2015} find stronger evolution of these populations with slopes $\alpha_{r_e}<-1$. \citet{mosleh2011} also finds steeper slopes for galaxies with UV bright selections. This hints at the fact that despite tracing part of a star-forming population, the LBG selection misses the global picture of star-forming galaxies that should be the dominant population at high redshifts.

There are also studies using galaxies selected by their photometric redshift \citep[e.g][]{trujillo2006,trujillo2007,buitrago2008,franx2008} and they find a global size evolution scaling with $\alpha_{r_e}=-0.40$, $-0.82$,$\sim-0.59$ at $0<z\lesssim3$, respectively, and also a stronger size evolution going from low to high mass galaxies. However, they have not specifically targeted the star-forming population at these redshifts.


\section{Conclusions}\label{section:conclusions}

We present the morphological characterization of SFGs selected through their H$\alpha$ luminosity and we compare their evolution across the last 11 Gyrs of the Universe ($z<2.23$). We compare the correlation of the structural parameters with the stellar mass and derive evolutionary trends for galaxy sizes (parameterized as the effective radius) and S\'ersic indices including the effect of cosmological dimming in the analysis. Our main results are:

\begin{itemize}
\item Cosmological surface brightness dimming (when using \textsc{GALFIT}) has a negligible impact ($\lesssim 10\%$) on the derived effective radius for all galaxies with $r_e<10$ kpc. We show that it impacts the derivation of the S\'ersic index, where we find a systematic underestimation of $\sim20\%$ at the higher redshifts in the artificially redshifted samples compared to the ones observed at $z\approx0$. This underestimation does not change the fact that the observed galaxies have surface brightness profiles similar to exponential disks.
\item  Effective radii of SFGs show moderate evolution ($\alpha_{r_e}\approx0.7\pm0.2$) from $z=2.23$ down to $z\approx0$. They have a range of galaxy sizes that grow from $\sim2$ kpc at $z=2.23$ up to $4-9$ kpc at $z\approx0$.
\item  We find that SFGs have typically disk-like profiles with a median value of $n\sim1.2$.
Their S\'ersic indices show negligible evolution ($\alpha_{n}$ is consistent with 0) across the same period with a median value of $1<n<1.5$, close to exponential disk profiles.
\item The stellar mass-size relation becomes steeper as we move towards lower redshifts. This hints at a stronger size evolution of the most massive SFGs when compared to the lower mass counterparts.
\end{itemize}

We use 4 different samples in the local Universe, 3 of them based on ongoing IFU surveys and one selected from SDSS as a control sample, to compare local to high redshift morphologies. Due to its diameter selected sample, the SFGs in the CALIFA survey are the most biased against a dynamic range in galaxy sizes. This limits the interpretations of the results if it is to be used as a local counterpart to the high-redshift samples being gathered with near-infrared IFU instruments. As for the SAMI and MaNGA samples, they seem to provide a representative morphological range of the local Universe, when compared to our NYU-VAGC control sample and, therefore, are more suitable for such comparisons.
 
Our results reveal that cosmological dimming plays a negligible role in the derivation of evolutionary trends on galaxy morphology for SFGs (and when using GALFIT). We show that SFGs grow in size, as seen from blue to UV rest-frame regions, by a factor of 2-3 since $z\sim2.23$ while their profile shapes remain the same (close to exponential disks). Interestingly, this growth is not observed in the same sample as seen from red-NIR regions \citep{stott2013b} and the observed differences are not due to sampling issues. This can also be linked to a scenario of inside-out star-formation as seen by \citet{nelson2015}. Although selection effects may play a role, it is possible that we are witnessing two distinct evolution paths for active star-forming regions and the underlying older stellar population across these redshifts. Alternatively, investigating new non-parametric size measurements, which fully account for cosmological surface brightness dimming \citep{ribeiro2016}, might provide new hints at size evolution trends. Moreover, our results put into perspective the galaxy morphologies of ongoing local IFU surveys and serve as a reference for future comparisons of local and high redshift IFU galaxy surveys.


\section*{Acknowledgements}
We thank the anonymous referee for the thorough review, positives comments, and constructive remarks which greatly improved the quality and clarity of this manuscript. We also thank J. Stott for sharing their size measurements that were used in the making of this paper.

This work was supported by Funda\c{c}\~{a}o para a Ci\^encia e a Tecnologia (FCT) through the research grant UID/FIS/04434/2013. APA, PhD::SPACE fellow, acknowledges support from the FCT through the felllowship PD/BD/52706/2014. DS acknowledges financial support from the Netherlands Organisation for Scientific research (NWO) through a Veni fellowship and from FCT through a FCT Investigator Starting Grant and Start-up Grant (IF/01154/2012/CP0189/CT0010).

Funding for SDSS-III has been provided by the Alfred P. Sloan Foundation, the Participating Institutions, the National Science Foundation, and the U.S. Department of Energy Office of Science. The SDSS-III web site is \url{http://www.sdss3.org/}.

SDSS-III is managed by the Astrophysical Research Consortium for the Participating Institutions of the SDSS-III Collaboration including the University of Arizona, the Brazilian Participation Group, Brookhaven National Laboratory, Carnegie Mellon University, University of Florida, the French Participation Group, the German Participation Group, Harvard University, the Instituto de Astrofisica de Canarias, the Michigan State/Notre Dame/JINA Participation Group, Johns Hopkins University, Lawrence Berkeley National Laboratory, Max Planck Institute for Astrophysics, Max Planck Institute for Extraterrestrial Physics, New Mexico State University, New York University, Ohio State University, Pennsylvania State University, University of Portsmouth, Princeton University, the Spanish Participation Group, University of Tokyo, University of Utah, Vanderbilt University, University of Virginia, University of Washington, and Yale University.

The HiZELS sample used in this work is publicly available from \citet{sobral2013}.

This work was only possible by the use of the following \textsc{python} packages: NumPy \& SciPy \citep{walt2011,jones2001}; Matplotlib \citep{hunter2007} and astropy \citep{robitaille2013}.


\bibliographystyle{mnras}
\bibliography{refs.bib}


\appendix

\section{Failure rate of convergence of \textsc{GALFIT}}\label{app:GALFIT_failures}

\begin{figure}
\centering
\includegraphics[width=\linewidth]{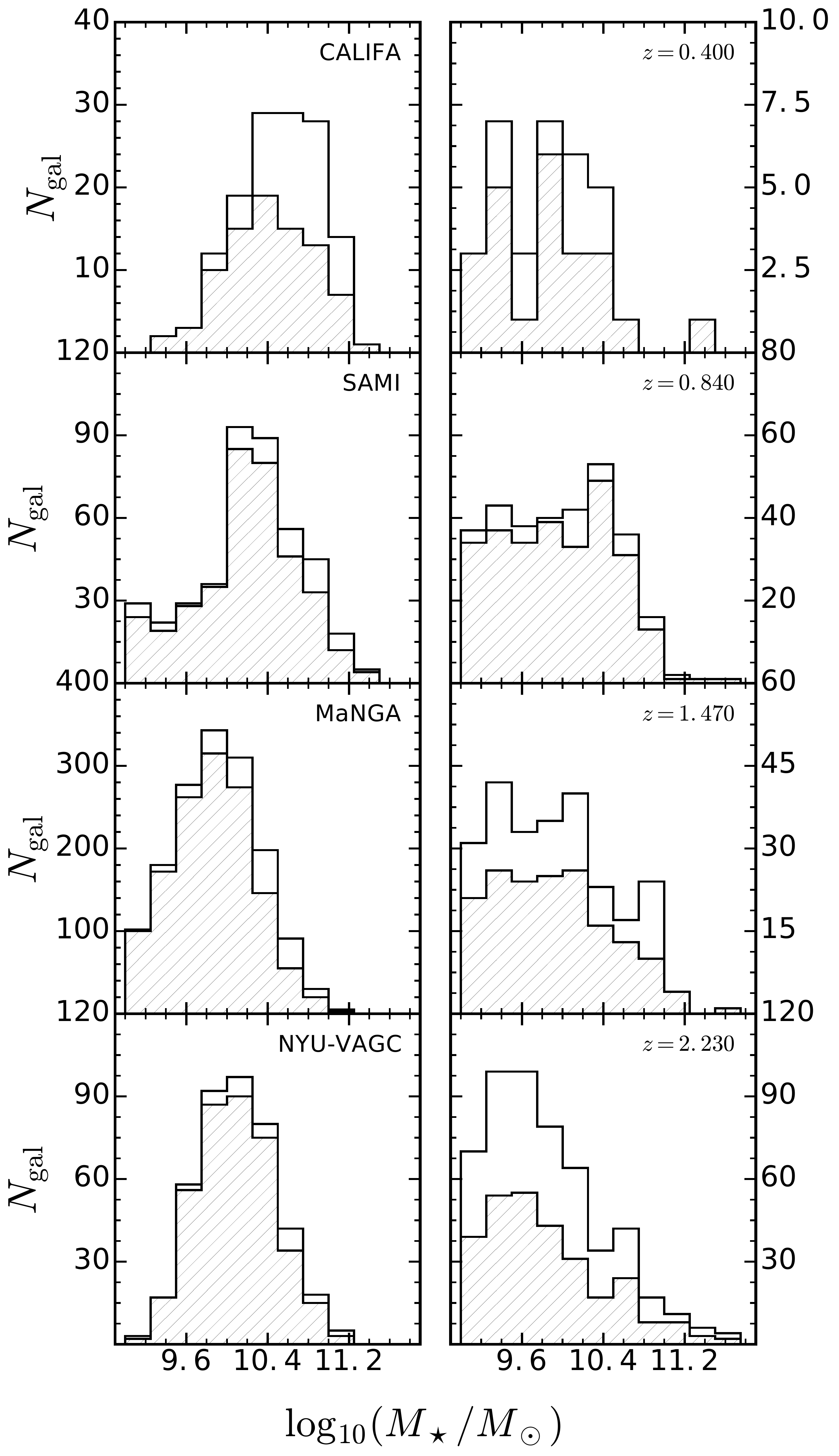}
\caption{The distribution of galaxies for which \textsc{GALFIT} converged (as diagonally hatched histogram) compared to the total sample (solid black line) that was analyzed as a function of their stellar mass.}
\label{fig:GALFIT_failed_mass}
\end{figure}

\begin{figure}
\centering
\includegraphics[width=\linewidth]{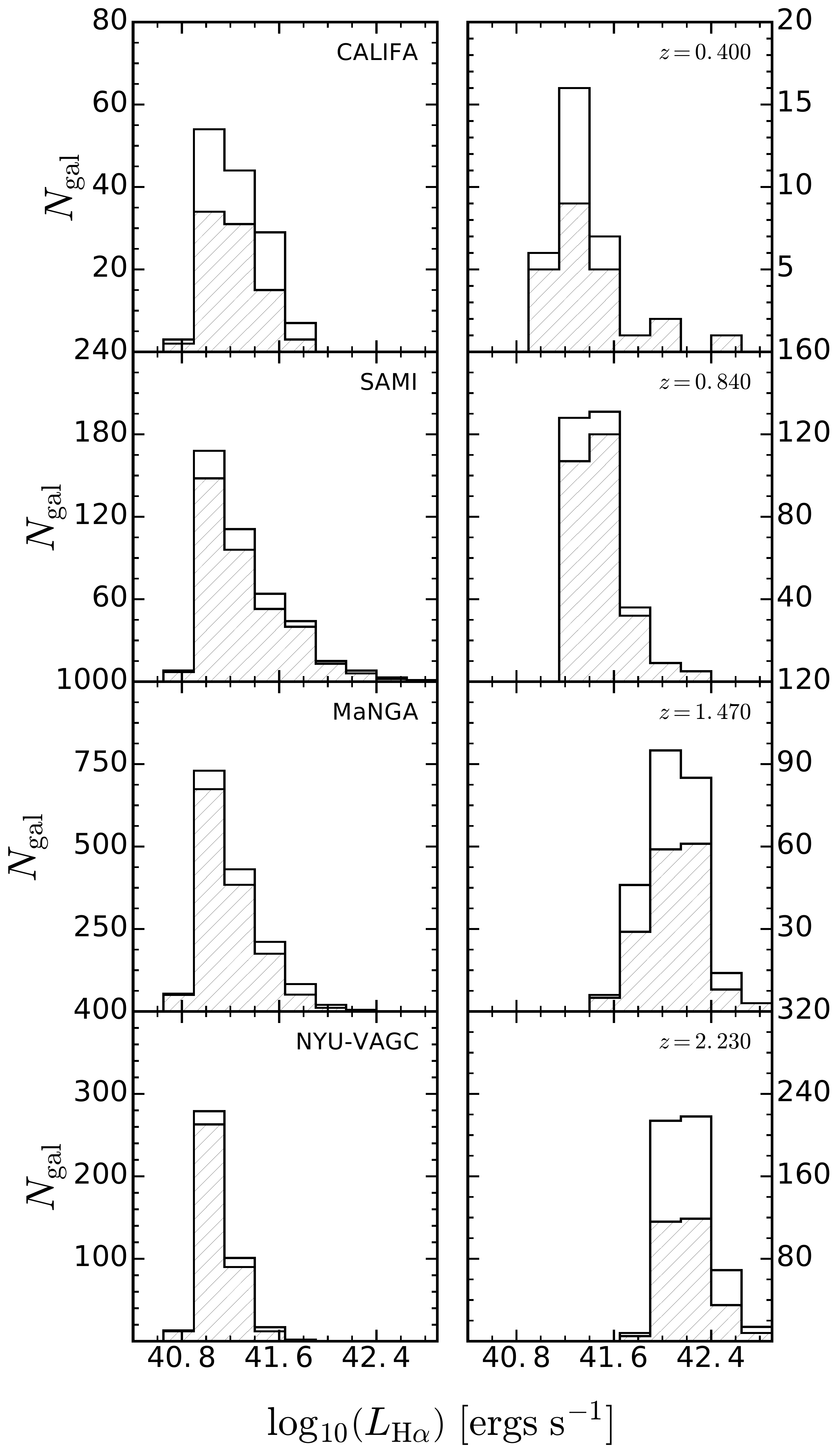}
\caption{The distribution of galaxies for which GALFIT converged (as diagonally hatched histogram) compared to the total sample (solid black line) that was analyzed as a function of their H$\alpha$ luminosity.}
\label{fig:GALFIT_failed_lumHa}
\end{figure}

We summarize in figures \ref{fig:GALFIT_failed_mass} and \ref{fig:GALFIT_failed_lumHa} the impact of the \textsc{GALFIT} failures to converge on the final stellar mass and H$\alpha$ luminosities distributions, respectively. It is evident that there is no bin in either in stellar mass or H$\alpha$ luminosity that is preferentially affected by a large failure rate when compared to other bins. This means that excluding galaxies for which \textsc{GALFIT} failed to converge from our final samples, from which we derive median structural parameters, does not introduce any additional bias in both stellar masses or H$\alpha$ luminosities.

\section{Further remarks on the impact of cosmological dimming}\label{app:others}

\begin{figure*}
\centering
\includegraphics[width=\linewidth]{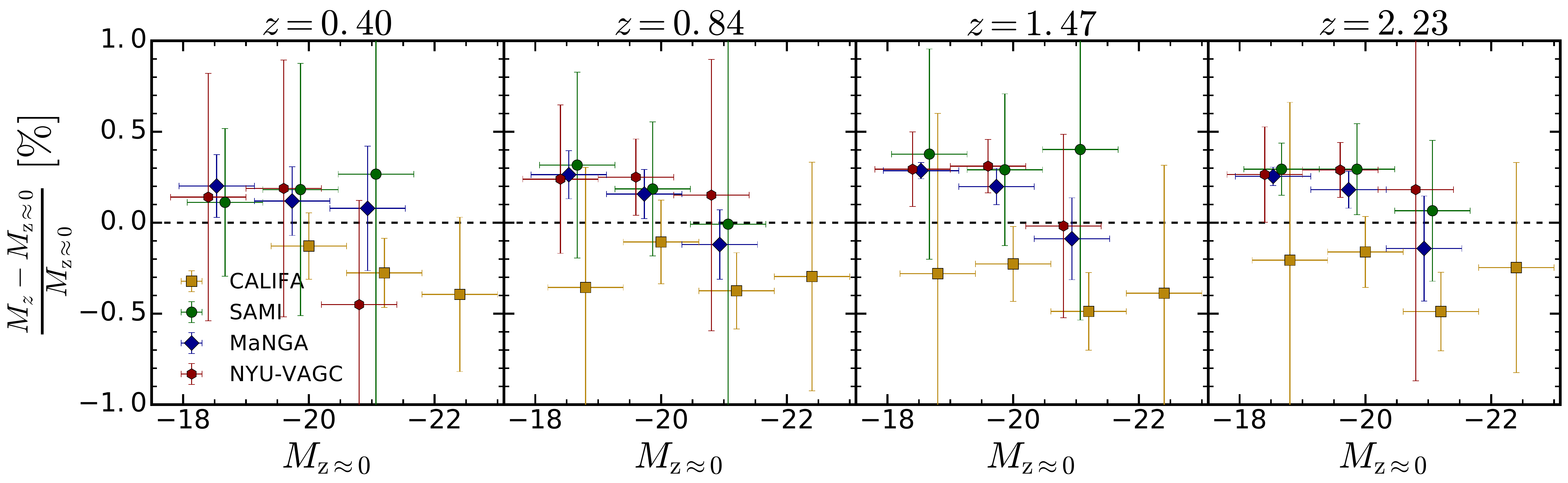}
\caption{Variable difference, $({M_{z}}-{M_{\mathrm{z\approx0}}})/{M_{\mathrm{z\approx0}}}$, for each of the local subsamples and each redshift (one per column). Each symbol represent the median values in bins (with a minimum of $10$ galaxies) of width of the horizontal error bar. The vertical error bar shows the error on the median value $\sigma/\sqrt{N_\mathrm{gal,bin}}$. The black dashed line pinpoints the accurate recovery. The recovered absolute magnitude is corrected for the luminosity evolution term of equation $\log_{10}\left[L(z)\right] = 0.45z+41.87$. The impact on this parameter is tiny ($\lesssim1\%$) at all redshifts. Small horizontal offsets were applied to improve the readability of the plot.}
\label{fig:results_mags_all_redshift_surveys}
\end{figure*}

\begin{figure*}
\centering
\includegraphics[width=\linewidth]{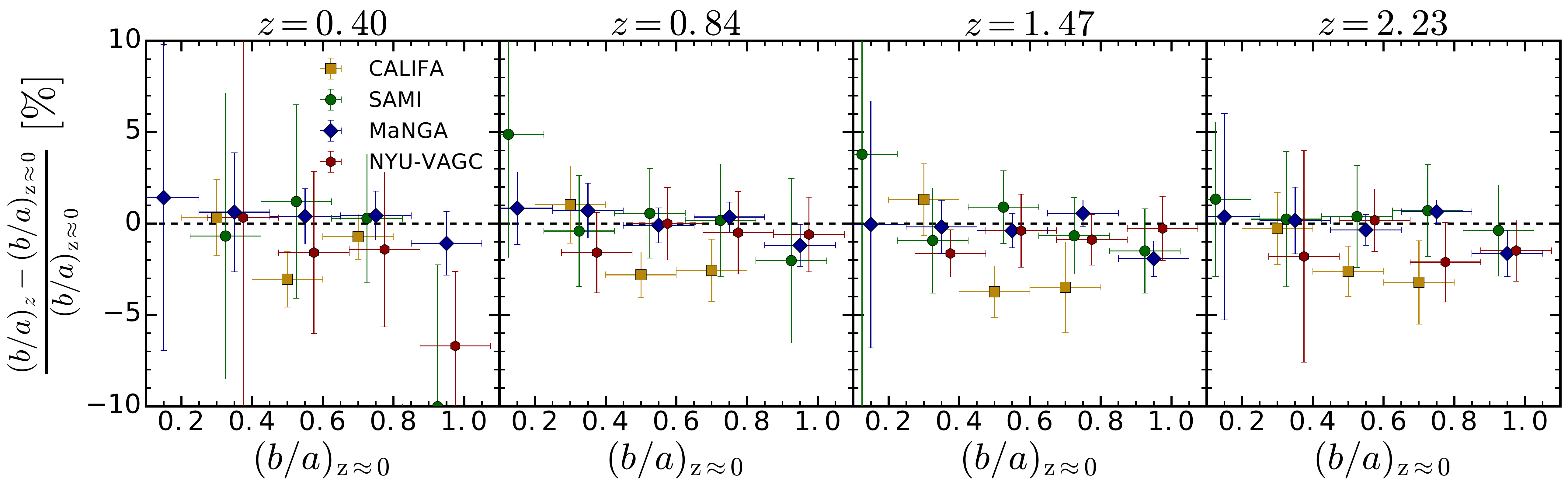}
\caption{Variable difference, $({(b/a)_{z}}-{(b/a)_{\mathrm{z\approx0}}})/{(b/a)_{\mathrm{z\approx0}}}$, for each of the local subsamples and each redshift (one per column). Each symbol represent the median values in bins (with a minimum of $10$ galaxies) of width of the horizontal error bar. The vertical error bar shows the error on the median value $\sigma/\sqrt{N_\mathrm{gal,bin}}$. The black dashed line pinpoints the accurate recovery. The impact on this parameter is small ($\lesssim5\%$) at all redshifts, with a slight tendency to overestimate at smaller $(b/a)$ and underestimate at higher $(b/a)$. Small horizontal offsets were applied to improve the readability of the plot.}
\label{fig:results_axrat_all_redshift_surveys}
\end{figure*}

We show in Figure \ref{fig:results_mags_all_redshift_surveys} the impact of the artificial redshift on the derived total magnitude of each galaxy. We compare absolute magnitudes and not the direct result from GALFIT, the apparent magnitudes, because it is the only way to compare magnitudes across different redshifts. We note that for comparison of the two quantities we correct the output absolute magnitude for the luminosity evolution that we impose for each redshift following the fit by \cite{sobral2013}. We show that the impact is close to zero ($<1\%$), which makes it the more stable parameter against cosmological dimming. As for the axis ratio of galaxies (Figure \ref{fig:results_axrat_all_redshift_surveys}) we recover, on average, the observed value at $z\approx0$, within an error of $\lesssim 5\%$. We observe a slight trend of overestimation of the axis ratio at smaller $(b/a)_{\mathrm{z\approx0}}$ and underestimation of the axis ratio at higher $(b/a)_{\mathrm{z\approx0}}$.

\begin{figure*}
\centering
\includegraphics[width=\linewidth]{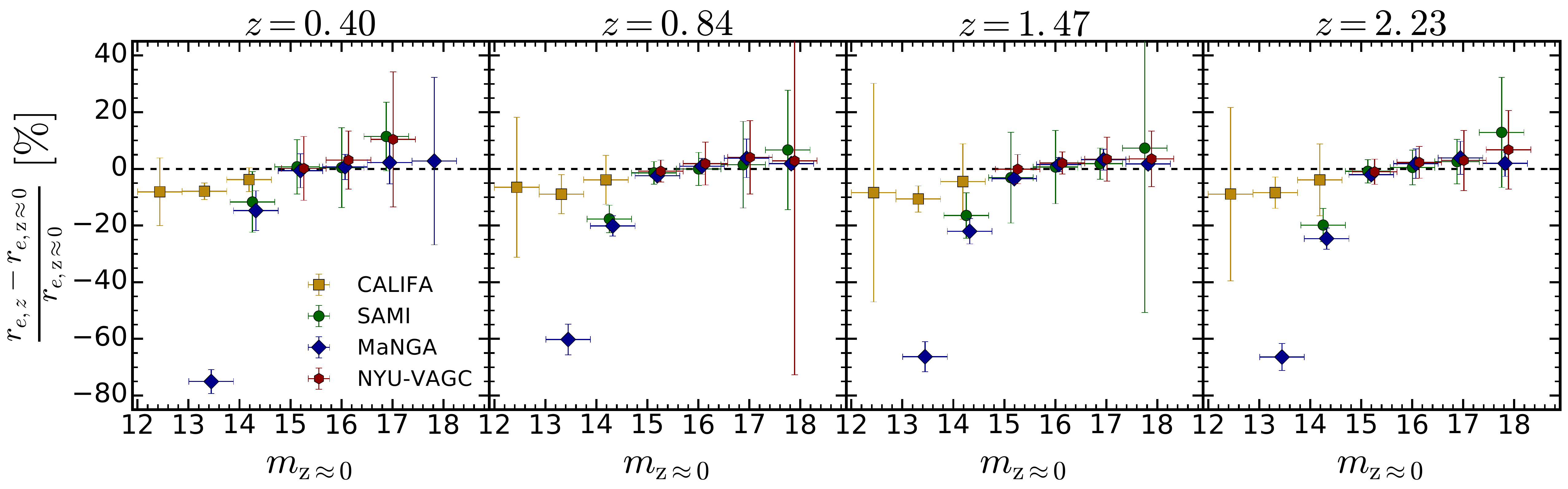}
\caption{The fraction difference between the recovered and input effective radius, $({r_{e,z}}-{r_{e,\mathrm{z\approx0}}})/{r_{e,\mathrm{z\approx0}}}$, for each of the local subsamples and each redshift (one per column) as a function of input magnitude. Each symbol represent the median values in bins (with a minimum of $10$ galaxies) of width of the horizontal error bar. The vertical error bar shows the error on the median value $\sigma/\sqrt{N_\mathrm{gal,bin}}$. The black dashed line pinpoints the accurate recovery. Small horizontal offsets were applied to improve the readability of the plot.}
\label{fig:results_reMAG_all_redshift_surveys}
\end{figure*}

We investigate further the impact of cosmological dimming by comparing our fractional errors on the effective radius and S\'ersic index to the input galaxy magnitude. In Figure \ref{fig:results_reMAG_all_redshift_surveys} we show that the brightest local galaxies are the ones most likely to have their sizes underestimated. The largest galaxies are the ones for which our sizes are not being recovered accurately as shown in Figure \ref{fig:results_rad_all_redshift_surveys} and they are also more likely to be the brightest galaxies in our sample which helps to explain partly our results. The size overestimation that we observe on the faint end is possibly explained due to the likelihood of local galaxies having a bulge+disk structure which can prevent a single S\'ersic profile to estimate the total extent of the galaxy. As we move the galaxy to higher redshifts the substructures tend to no-longer be resolved by the instrument PSF and GALFIT can more successfully measure sizes for the entire galaxy, hence estimating a larger size than the one measured at low redshift for the same galaxy. On the other hand, if the bulge-to-total light ratio is large, it may imply that we completely loose the disk component of the galaxy and end up underestimating the size of the galaxy. This latter effect is expected to happen on the brighter galaxies since those are the ones we expected to have more likely experienced at least one major merger which induces the formation of a prominent central bulge.

\begin{figure*}
\centering
\includegraphics[width=\linewidth]{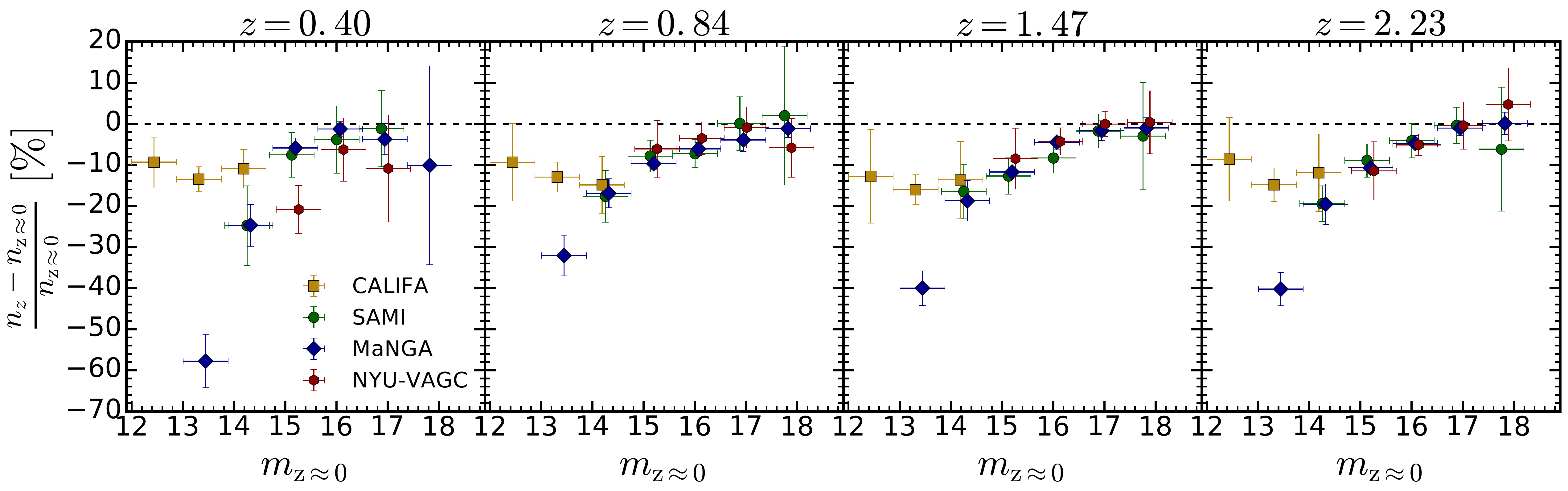}
\caption{The fraction difference between the recovered and input S\'ersic index, $({n_{z}}-{n_{\mathrm{z\approx0}}})/{n_{\mathrm{z\approx0}}}$, for each of the local subsamples and each redshift (one per column) as a function of input magnitude. Each symbol represent the median values in bins (with a minimum of $10$ galaxies) of width of the horizontal error bar. The vertical error bar shows the error on the median value $\sigma/\sqrt{N_\mathrm{gal,bin}}$. The black dashed line pinpoints the accurate recovery. Small horizontal offsets were applied to improve the readability of the plot.}
\label{fig:results_nMAG_all_redshift_surveys}
\end{figure*}

Since both effective parameters are linked through the same equation, we expect that a failure to reproduce the original effective radius leads in turn to a large error on the S\'ersic index of the corresponding profile. And since the S\'ersic index is the most unstable parameter of the profile we are likely witnessing in Figure \ref{fig:results_nMAG_all_redshift_surveys} a simple consequence of the results shown before for the effective radius. 

We have further separated the sample in two axis ratio bins (below and above $b/a=0.5$) and re-did Figures \ref{fig:results_nMAG_all_redshift_surveys} and \ref{fig:results_nMAG_all_redshift_surveys}. The results we find in these case are qualitatively the same and so we conclude that the axis ratio as no major impact on our ability to recover sizes and S\'ersic indexes.

\begin{figure*}
\centering
\includegraphics[width=\linewidth]{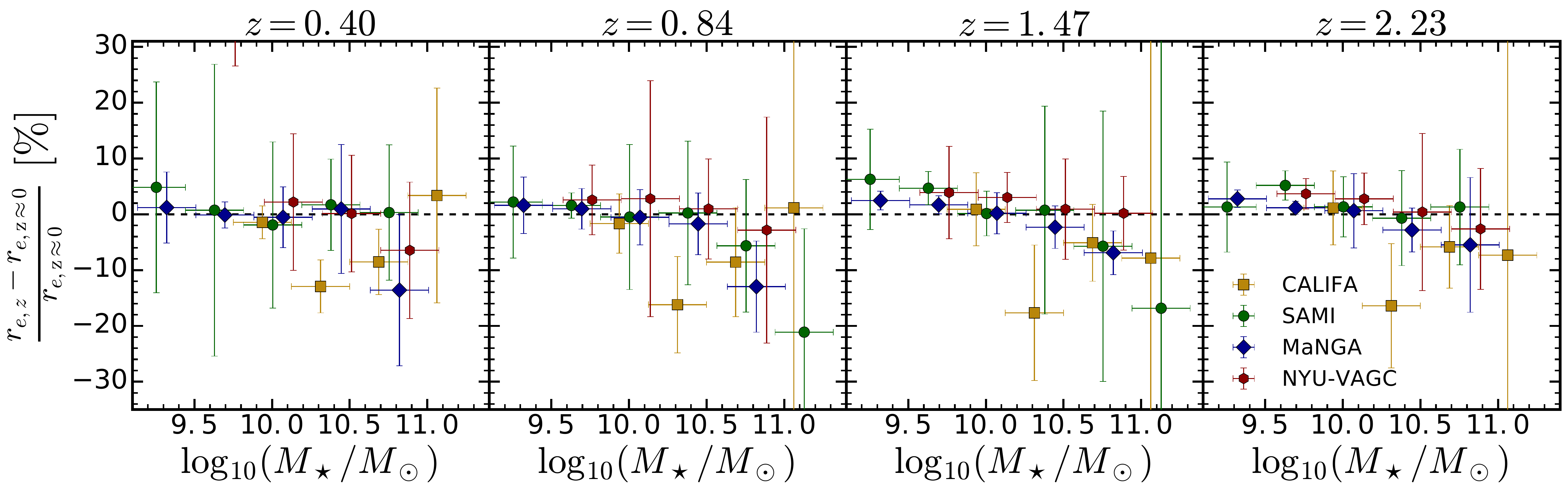}
\caption{The fraction difference between the recovered and input effective radius, $({r_{e,z}}-{r_{e,\mathrm{z\approx0}}})/{r_{e,\mathrm{z\approx0}}}$, for each of the local subsamples and each redshift (one per column) as a function of input stellar mass. Each symbol represent the median values in bins (with a minimum of $10$ galaxies) of width of the horizontal error bar. The vertical error bar shows the error on the median value. The black dashed line pinpoints a fractional difference of 0, i.e., when the recovered effective radius is the same as input.  We do not observe a strong trend of fractional error deviation as a function of stellar mass. Nonetheless, the most massive galaxies are expected to have their sizes underestimated at a level of $\lesssim20\%$. Small horizontal offsets were applied to improve the readability of the plot. On the $z=0.40$ panel there is one point from the SAMI survey at $({r_{e,z}}-{r_{e,\mathrm{z\approx0}}})/{r_{e,\mathrm{z\approx0}}}\sim40\%$ for $\log_{10}(M_\star/M_\odot)\sim9.75$ that was excluded for visualization purposes.}
\label{fig:results_radius_vs_mass_all_redshift_surveys}
\end{figure*}

\bsp
\label{lastpage}
\end{document}